\documentclass[fleqn,usenatbib]{mnras}

\usepackage{newtxtext,newtxmath}
\usepackage{soul}

\usepackage[T1]{fontenc}

\DeclareRobustCommand{\VAN}[3]{#2}
\let\VANthebibliography\thebibliography
\def\thebibliography{\DeclareRobustCommand{\VAN}[3]{##3}\VANthebibliography}

\usepackage{graphicx}	
\usepackage{amsmath}	

\usepackage{bbm}
\usepackage{cleveref}
\usepackage{multirow}
\usepackage{pgfplots}
\usepackage{placeins}
\usepackage{tikz}
\usepackage{upgreek}
\usepackage{booktabs}

\definecolor{ccqqqq}{rgb}{0.8,0,0}
\graphicspath{./Figures/}

\pgfplotsset{compat=1.15}

\usetikzlibrary{arrows}

\crefname{equation}{equation}{equations}
\Crefname{equation}{Equation}{Equations}
\crefname{figure}{Fig.}{Figs.}
\Crefname{figure}{Fig.}{Figs.}
\crefname{table}{Table}{Tables}
\Crefname{table}{Table}{Tables}
\crefname{section}{Section}{Sections}
\Crefname{section}{Section}{Sections}

\newcommand{\vs}{\mathbfit{s}}
\newcommand{\vr}{\mathbfit{r}}
\newcommand{\vvel}{\mathbfit{v}}
\newcommand{\vq}{\mathbfit{q}}

\newcommand{\vhz}{\hat{\mathbfit{z}}}

\newcommand{\Ae}{AE}
\newcommand{\hmpc}{h^{-1}\,{\rm Mpc}}

\newcommand{\hmpcinv}{h\,{\rm Mpc}^{-1}}

\newcommand{\hmpcdens}{h^3\,{\rm Mpc}^{-3}}

\newcommand{\kms}{{\rm km}\,{\rm s}^{-1}}

\newcommand{\imagetop}[1]{\vtop{\null\hbox{#1}}}

\newcommand{\vect}[1]{\mathbfit{#1}}
\DeclareMathOperator*{\argmin}{arg\,min}

\newcommand{\inp}{I}
\newcommand{\targ}{T}
\newcommand{\est}{\hat{\targ}}
\newcommand{\vinp}{\mathbfit{I}}
\newcommand{\vtarg}{\mathbfit{T}}
\newcommand{\vest}{\hat{\vtarg}}
\newcommand{\vb}{\mathbfit{b}}
\newcommand{\mw}{\mathbf{w}}
\newcommand{\mC}{\mathbf{C}}
\newcommand{\mse}{\mathrm{MSE}}

\newcommand{\wf}{\mathrm{WF}}
\newcommand{\lin}{\mathrm{lin}}

\title[Neural network field reconstruction]{Large-scale density and velocity field reconstructions with neural networks}

\author[P.~Ganeshaiah Veena, R.~Lilow and A.~Nusser]{
Punyakoti Ganeshaiah Veena,$^{1}$\thanks{E-mail: \href{mailto:punyakoti.g@campus.technion.ac.il}{punyakoti.g@campus.technion.ac.il}}
Robert Lilow$^{1}$\thanks{E-mail: \href{mailto:rlilow@campus.technion.ac.il}{rlilow@campus.technion.ac.il}}
and Adi Nusser$^{1}$\thanks{E-mail: \href{mailto:adi@physics.technion.ac.il}{adi@physics.technion.ac.il}}
\\
$^{1}$Department of Physics, Technion, Haifa 3200003, Israel
}

\date{Accepted XXX. Received YYY; in original form ZZZ}

\pubyear{2022}

\begin{document}
\label{firstpage}
\pagerange{\pageref{firstpage}--\pageref{lastpage}}
\maketitle


\begin{abstract}
We assess a neural network (NN) method for reconstructing 3D cosmological density and velocity fields (target) from discrete and incomplete galaxy distributions (input). We employ second-order Lagrangian Perturbation Theory to generate  a large ensemble of mock  data to train an autoencoder (AE) architecture with a Mean Squared Error (MSE) loss function. The AE successfully captures nonlinear features arising from gravitational dynamics and the discreteness of the galaxy distribution. It preserves the positivity of the reconstructed density field and exhibits a weaker suppression of the power on small scales than the traditional linear Wiener filter (WF), which we use as a benchmark. In the density reconstruction, the reduction of the AE MSE relative to the WF is $\sim 15 \%$, whereas for the velocity reconstruction a relative reduction of up to a factor of two can be achieved. The \Ae\  is  advantageous  to the WF at recovering the distribution of the target fields, especially at  the tails.
In fact, trained with an MSE loss, any  NN estimate approaches
the unbiased mean of the underlying target given the
input. This implies a  slope of unity in the linear regression of the true on the NN-reconstructed field. 
Only for the special case of Gaussian fields, the NN and WF estimates are equivalent.
Nonetheless,  we  also recover a linear  regression slope of unity  for the WF with non-Gaussian fields.
\end{abstract}




\section{Introduction}

Standard cosmology relies on gravity as the main driver for the amplification of tiny initial density fluctuations into the observed variety of structure in the late-time Universe.
The basic theory is appealing in its simplicity.
Given a set of appropriate initial conditions, the subsequent evolution of 
mass density fluctuations 
depends in a simple manner on only a few global parameters. These include the mean mass densities of baryons and dark matter (DM), and  the mean energy density of a (dark energy) component that 
provides the mechanism for accelerating the expansion of the  cosmological  background. 

Therefore, observations of the  large-scale ($\gtrsim $ a few Mpc) distribution and motion of galaxies encode signatures of the cosmic energy density content and nature of gravity on cosmic scales.
Two  categories of observational data are relevant to the current study. The first is surveys of galaxy positions and redshifts \citep[e.g.][]{huchra_survey_1983,neugebauer_infrared_1984,colless_2df_2001,jones_6df_2009,lavaux_2m++_2011,huchra_2mass_2012,abazajian_seventh_2009,ahn_tenth_2014,macri_2mass_2019,ahumada_16th_2020},  
and the second is catalogs of distance indicators from which the peculiar velocities of galaxies (deviations from a pure Hubble flow) can be inferred \citep[e.g.][]{a82,Lynden-Bell1988,strauss_density_1995,spring07,turnbull_cosmic_2012, springob_6df_2014, hong_2mtf_2019, stahl_peculiar-velocity_2021, tully_cosmicflows-4_2022}. Galaxy redshifts are relatively easy to obtain and thus redshift surveys have been  the main  probe of  cosmological models. 
Peculiar velocities contain complementary information to redshift surveys, and beyond that are directly proportional to the gravitational force field independent of any bias between the galaxy distribution and the underlying mass density field.
We emphasize, though, that  peculiar velocities are harder to obtain and hence the corresponding data sets are significantly smaller than redshift surveys. 

Since a galaxy redshift differs from its distance by the line-of-sight  peculiar velocity, the appearance of  structure in redshift space 
deviates from the actual structure in real space. The deviations (distortions) are  systematic due to the coherence between peculiar motions and 
the density of matter, as predicted by 
gravitational instability  \citep[e.g.][]{Peeb80,nussetal91}. These redshift-space distortions (RSD)
 \citep{Davis1983}  break the statistical isotropy of structure 
and offer a powerful tool of constraining cosmological parameters via the analysis of correlation functions \citep[e.g.][]{kaiser_clustering_1987,percival_2df_2004,blake_wigglez_2011,beutler_6df_2012,samushia_interpreting_2012,howlett_clustering_2015,achitouv_consistency_2017,alam_clustering_2017,blake_power_2018,bautista_completed_2020,gil-marin_completed_2020,tamone_completed_2020,de_mattia_completed_2020}. 

The  distribution of galaxies in redshift surveys can in itself be used to reconstruct  the corresponding peculiar velocity field. The reconstruction  relies on the tight coherence between density and velocity as implied by gravitational instability and requires an assumption of the cosmological parameters. 
The reconstructed velocity can then be employed in several ways. It can 
be used to recover distances of galaxies  from their redshifts  and subsequently
obtain the real-space density field of galaxies. Another application is the galaxy luminosity modulation method \citep{1980ApJ...242..448Y,branchini2012,Feix2015}, where the cosmological parameters are tuned to  minimize  scatter in the galaxy luminosities inferred from the fluxes and recovered distances. 
The reconstructed velocities can be compared with the directly observed velocities from catalogs of distance indicators. This highly important  comparison has been performed using a variety of galaxy redshift surveys and peculiar velocity data \citep[e.g.][]{kaiser_large-scale_1991,hudson_optical_1994,shaya_action_1995,DNW96,pike_cosmological_2005,davis_local_2011,ma_comparison_2012,turnbull_cosmic_2012,carrick_cosmological_2015,boruah_cosmic_2020,said_joint_2020,lilow_constrained_2021,stahl_peculiar-velocity_2021}. The comparison so far has yielded important constraints on the gravitational instability paradigm for structure formation as well as the biasing relation between galaxies and the underlying mass density field (mainly due to the DM).

Observational data  are  usually noisy, patchy and sparse. It has been a constant struggle to make most of  the incomplete data to infer accurate information. 
In the current work we focus on galaxy redshift surveys, with the goal of 
providing estimates for the underlying density and velocity fields from the incomplete and sparsely sampled  galaxies in  these surveys. 

Several reconstruction methods have been proposed  in the literature. A 
broad class of applications  involve two main ingredients:  smoothing of the galaxy redshift survey, to provide a galaxy density field in redshift space, and the dynamical   relation between the velocity and density fields.  The main purpose of  smoothing is  to mitigate  the discreteness noise (shot noise) in the galaxy distribution. The smoothing kernel could be a Gaussian with a variable  width designed  to match the selection function of the data  \citep[e.g.][]{yahil_redshift_1991}.
Another strategy is to apply a Wiener filter (WF) \citep{wiener_extrapolation_1949} to the galaxy distribution to infer an underlying field \citep[e.g.][]{zaroubi_wiener_1995,fisher_wiener_1995,webster_wiener_1997,schmoldt_density_1999,erdogdu_reconstructed_2006,lilow_constrained_2021}.
Given an observed field, the WF estimate is the linear solution that minimises the variance of the difference between true and estimated underlying fields.
The expression of the variance is written in the ensemble average sense 
and hence requires the auto-correlation functions of the true and observed fields, respectively, as well as their cross correlations.
The WF estimate has the nice property that for Gaussian true and observed fields, it is identical to the mean, {median} and maximum of the posterior probability, i.e.~the probability distribution function (PDF) of the true field given the observations. 

A conceptually different approach is based on machine learning. 
Machine learning describes any type of algorithm that attempts to optimize a certain task based on   a suitable training data set\footnote{Machine learning encompasses a vast variety of algorithms, and we refer the reader to \citet{bishop_pattern_2006} and \citet{goodfellow_deep_2016} for a comprehensive overview of this field. 
}. In this work, we employ a neural network (NN) for the task of recovering the underlying (true) cosmological density or velocity fields from  the observed discrete distribution of mass tracers in galaxy redshift surveys.
The relevant training data set consists of mock underlying fields and the mock observations extracted from those.

A NN does not require the explicit evaluation of covariance matrices and  PDFs, which  is a major bottleneck in 
standard inference  via maximum likelihood/posterior  techniques. 
Furthermore, there is no need for any assumption on the form of the relationship between the observed and underlying fields.
NNs can incorporate almost any nonlinear and nonlocal relationship that exists between those. 

The importance of NNs in astronomy has long been recognized, for example, for galaxy morphology classification \citep[e.g.][]{folkes_artificial_1996,lahav_neural_1996,ball_galaxy_2004}, for object detection from images \citep[e.g.][]{bertin_sextractor_1996,andreon_wide_2002} and for photometric redshift estimation \citep[e.g.][]{firth_estimating_2003,vanzella_photometric_2004,collister_annz_2004}.
In recent years,  NNs  have become  popular in studies of large-scale structure reconstructions. Examples include the reconstruction of convergence maps from weak lensing data \citep{Niall_Deep_learning_2020}, the reconstruction of the matter density field from observed galaxy distribution and line-of-sight velocities \citep{hong_revealing_2021}, the reconstruction of the matter velocity field from a matter density field \citep{wu_cosmic_2021}, the reconstruction of the initial matter density field from an evolved matter density field \citep{shallue_reconstructing_2022}, the reconstruction of individual cluster velocities from their surrounding galaxy distribution to measure the kinetic Sunyaev-Zel’dovich effect \citep{Tanimura_convolutional_2022}, and the improvement of a traditional maximum a posteriori density field reconstruction by learning a non-Gaussian density prior \citep{rouhiainen_-noising_2022}.

Our application is closest to \citet{hong_revealing_2021}, but has some key differences: \citet{hong_revealing_2021} considers the reconstruction of the density field given the observed galaxy positions and radial velocities from a sparse peculiar velocity survey, like Cosmicflows-3 \citep{tully_cosmicflows-3_2016}, whereas we consider the reconstruction of both density and velocity fields given only the observed galaxy positions from a dense redshift survey, like the 2MASS Redshift Survey \citep{huchra_2mass_2012}.
Furthermore, our work focuses more on the interpretation of NNs, as we lay out in the following.

The main criticism of NNs is related to the interpretability of the inferred mapping  between observed and underlying fields.
Therefore, one of the main goals of this paper is to elucidate the connection between NNs  and well-known statistical estimators, namely the mean posterior estimator and the popular WF \citep{goodfellow_deep_2016}. For Gaussian fields, we show that a NN minimizing the commonly adopted Mean Squared Error (MSE) loss function is identical to the WF. Furthermore, we employ 2\textsuperscript{nd}-order Lagrangian perturbation theory (2LPT) \citep{moutarde_precollapse_1991,bouchet_weakly_1992,Buchert1993,Gramann1993,bouchet_perturbative_1995,Zheligovsky2014} to efficiently generate non-Gaussian fields. We explore the performance of the NN reconstruction on these non-Gaussian fields and compare it to the performance of the WF, to demonstrate that the NN reconstruction approaches the nonlinear mean posterior estimate, which captures nonlinear features and reduces the MSE beyond the capability of the linear WF reconstruction. Additionally, the traditional WF acts as a benchmark for assessing the NN method.

In \cref{sec:methodology}, we describe the NN methodology for the reconstruction of  density and velocity fields. The section also clarifies the connection with the WF and contains details of our specific NN architecture.
\Cref{sec:data} presents the Gaussian and 2LPT mock data, generated from a realistic power spectrum, for training the NN. 
In \cref{sec:results}, the NN and WF are applied to  the mock fields. Here the equivalence of NN and WF for Gaussian data is explicitly demonstrated.
Afterwards, a detailed assessment of the  performance of the NN reconstruction is presented for the 2LPT mocks.
In \cref{sec:nbar}, we explore the dependence of the MSE on the mean galaxy density.
We conclude  in \cref{sec:discussion} with a discussion of the main results and potential improvements and  extensions.
The \cref{sec:2LPT_data_generation} provides technical details on 2LPT and the corresponding data generation. In \cref{sec:WF_computation}, we describe the numerical computation of the WF.

\section{Neural network methodology}
\label{sec:methodology}
Given an observed field on a grid as input, our goal is to infer an estimate of an underlying target field associated with the observed field. In our study, the input is a noisy cosmological density field and the target can be the underlying matter density field and/or the peculiar velocity associated with the observed density. We use a NN to design a mapping between input and target fields by extracting connections and correlations between an ensemble of mock observed and their corresponding mock underlying fields, which together form the training data set.

More specifically, we are employing a feedforward NN \citep{goodfellow_deep_2016}. In such a NN, the desired mapping is modelled by means of a directed acyclic network of nodes connecting the observed field values (input layer) to the estimated underlying field values (output layer), typically arranged in a sequence of intermediate hidden layers. The number of hidden layers determines the depth of the NN, and the number of nodes per hidden layer determines its width. The value of the $i$-th node in the $n$-th layer, $\mathcal{N}^{(n)}_i$, is obtained by computing the linear combination of the previous layer's node values, $\mathcal{N}^{(n-1)}_j$, usually followed by the application of  some nonlinear activation function $A^{(n)}_i$,
\begin{equation}
    \mathcal{N}^{(n)}_i = A^{(n)}_i\biggl( \sum_{j} \, w^{(n,n-1)}_{ij} \, \mathcal{N}^{(n-1)}_j + b^{(n)}_i \biggr) \,,
    \label{eq:neural_net_node_value}
\end{equation}
with weight parameters $w^{(n,n-1)}_{ij}$ and bias parameters $b^{(n)}_i$. Given an input field, the mapping to the target field estimate is then evaluated by forward-propagating the node values from the input to the output layer by iteratively applying \cref{eq:neural_net_node_value}.

During the training of the NN, the free weight and bias parameters are optimised by minimising a given loss function. The most common optimization methods are variants of (mini-batch) stochastic gradient descent \citep{goodfellow_deep_2016}. In this method, the training data set is split into separate mini-batches, which are then processed one at a time. For each mini-batch the average gradient of the loss function is computed and used to update all parameters by back-propagating through the NN. The next mini-batch is then processed using the updated parameters. Once all mini-batches are processed, this concludes a single \emph{training epoch}. This process is repeated for a fixed number of epochs or until some convergence criterion is satisfied.

The properties of the mapping that a given NN learns is defined by two ingredients: (i) The chosen architecture of the NN (network graph structure and activation functions) constrains the available function space for the mapping from input to target fields. (ii) The optimal mapping in this available space is then defined as the minimum of the chosen loss function. For a given choice of these ingredients, we can rigorously infer the statistical interpretation of the resulting estimator.

\subsection{Interpretation of MSE loss function}
\label{sec:methodology:MSE}
Consider a training data set consisting of mock input fields $\vinp^\alpha$ and corresponding mock target fields $\vtarg^\alpha$, where $\alpha = 1,\dotsc,M$ labels the individual field realizations. The field values in the $j$-th grid cell are denoted by $\inp^\alpha_j$ and $\targ^\alpha_j$, $j = 1,\dotsc,N$. In our application, $\vinp^\alpha$ are noisy observed density fields, and $\vtarg^\alpha$ are either the underlying true density or peculiar velocity fields. For a given NN architecture $\est[\boldsymbol{\lambda}]$, parameterised by the vector $\boldsymbol{\lambda}$ representing the weights and biases. We define the estimator $\vest$ by minimising an appropriate  loss function $L(\vest[\boldsymbol{\lambda}])$,
\begin{equation}
    \vest = \vest[\hat{\boldsymbol{\lambda}}] \,, \quad \hat{\boldsymbol{\lambda}} = \argmin_{\boldsymbol{\lambda}} L(\vest[\boldsymbol{\lambda}]) \,.
\end{equation}
Note that no assumption is made on the relation between input and target fields. It can be nonlinear as well as nonlocal; as is the case, for example, when we reconstruct the true velocity field from the noisy observed densities.

In this work we consider the commonly adopted Mean Squared Error (MSE) loss function,
\begin{equation}
    L^\mse(\vest[\boldsymbol{\lambda}]) = \frac{1}{M N} \sum_{\alpha=1}^M \sum_{j=1}^N \, \bigl(\targ^\alpha_j - \est_j[\boldsymbol{\lambda}](\vinp^\alpha)\bigr)^2 \,.
    \label{eq:MSE_loss}
\end{equation}
To re-derive the well-known statistical interpretation of the resulting minimum MSE estimator $\vest^\mse$, we consider the limit of an infinitely large set of training fields, allowing us to rewrite $L^\mse$ as a sampling over the joint distribution of input and target fields,
\begin{align}
    L^\mse(\vest[\boldsymbol{\lambda}]) &\xrightarrow{M \rightarrow \infty} \frac{1}{N} \sum_{\vinp,\vtarg} P(\vinp,\vtarg) \sum_{j=1}^N \, \bigl(\targ_j - \est_j[\boldsymbol{\boldsymbol{\lambda}}](\vinp)\bigr)^2 \\
    &= \frac{1}{N} \sum_\vinp P(\vinp) \sum_\vtarg P(\vtarg|\vinp) \sum_{j=1}^N \, \bigl(\targ_j - \est_j[\boldsymbol{\lambda}](\vinp)\bigr)^2 \,.
     \label{eq:MSE_loss_via_posterior}
\end{align}
In the second line we used Bayes' theorem, to split the joint distribution into the evidence $P(\vinp)$ and the posterior $P(\vtarg|\vinp)$. The minimum MSE estimator is defined by setting the first variation of $L^\mse$ w.r.t.~$\vest[\boldsymbol{\lambda}]$ to zero,
\begin{align}
    0 &= \frac{\updelta L^\mse(\vest[\boldsymbol{\lambda}])}{\updelta \est_i[\boldsymbol{\lambda}]} \, \biggr|_{\vest[\boldsymbol{\lambda}]=\vest^\mse} \\
    &= -\frac{2}{N} \, P(\vinp) \, \sum_\vtarg \, P(\vtarg|\vinp) \, \bigl(\targ_i - \est^\mse_i(\vinp)\bigr) \,.
\end{align}
It is thus exactly given by the mean of the posterior distribution,
\begin{equation}
    \est^\mse_i(\vinp) = \sum_\vtarg \, P(\vtarg|\vinp) \, \targ_i \, = \langle \targ_i| \vinp \rangle,
    \label{eq:mean_posterior_estimator} 
\end{equation}
i.e.~the ensemble average of the target field $\vtarg$ conditional on the given input field $\vinp$ \citep{goodfellow_deep_2016}.\footnote{Similarly, the Mean Absolute Error loss function yields the median of the posterior \citep{goodfellow_deep_2016}.} In our application this corresponds to the mean underlying density or velocity field conditional on the observations.

Note that a NN with MSE loss function can only approach the mean posterior estimator given in \cref{eq:mean_posterior_estimator} if (i) the training set is sufficiently large to approximate the statistical mean of the underlying posterior probability distribution by averaging over the training set, and (ii) if the NN architecture is sufficiently complex to express the mapping from the input to the mean of the posterior.\footnote{It may still  approach a local rather than the global minimum of the MSE loss function. However, the random fluctuations in (mini-batch) stochastic gradient descent algorithms used for the  minimization reduce the likelihood of settling  in local minima and thus help approaching the global minimum \citep{goodfellow_deep_2016}.} We can demonstrate the effect of breaking the second condition by considering a simple linear NN architecture, as demonstrated in the following.

\begin{figure}
    \centering
    \begin{tikzpicture}[scale=0.5, x=1cm, y=1cm, line width=1pt]
	\draw (0,5)-- (7,5);
	\draw (0,3)-- (7,5);
	\draw (0,1)-- (7,5);
	\draw (0,-3)-- (7,5);
	\draw (0,5)-- (7,3);
	\draw (0,3)-- (7,3);
	\draw (0,1)-- (7,3);
	\draw (0,-3)-- (7,3);
	\draw (0,1)-- (7,1);
	\draw (0,-3)-- (7,1);
	\draw (0,5)-- (7,-3);
	\draw (0,3)-- (7,-3);
	\draw (0,1)-- (7,-3);
	\draw (0,-3)-- (7,-3);
	\draw (0,3)-- (7,1);
	\draw (0,5)-- (7,1);
	\draw [fill=ccqqqq] (0,1) circle (7pt);
	\draw [fill=ccqqqq] (0,5) circle (7pt);
	\draw [fill=ccqqqq] (0,3) circle (7pt);
	\draw [fill=black] (0,-0.5) circle (1pt);
	\draw [fill=black] (0,-1) circle (1pt);
	\draw [fill=black] (0,-1.5) circle (1pt);
	\draw [fill=ccqqqq] (0,-3) circle (7pt) ; 
	\draw [fill=ccqqqq] (7,5) circle (7pt);
	\draw [fill=ccqqqq] (7,3) circle (7pt);
	\draw [fill=ccqqqq] (7,1) circle (7pt);
	\draw [fill=ccqqqq] (7,-3) circle (7pt);
	\draw [fill=black] (7,-0.5) circle (1pt);
	\draw [fill=black] (7,-1) circle (1pt);
	\draw [fill=black] (7,-1.5) circle (1pt);
	\node[left] at (-0.5,0.8) {\large $\inp_j$} ;
	\node[right] at (7.5,0.8) {\large $\hat{\targ}^\wf_i$} ;
	\node[below] at (3.5,-3) {\large $w_{ij}, b_i$} ;
\end{tikzpicture}
    \caption{Schematic of a densely connected linear NN with only an input and an output layer. Every  node (red circles) $j$ in the input layer is connected to all  nodes $i$ in the output layer via weights $w_{ij}$ and biases $b_i$. This network does not have any hidden layers or {nonlinear} activations.  Such a  network is equivalent to a Wiener filter when optimised on a MSE loss (cf.~\cref{sec:methodology:wiener_filter}).}
    \label{fig:ZHL}
\end{figure}
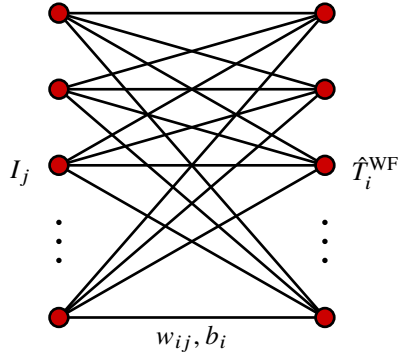
\subsection{Wiener filter via neural network}
\label{sec:methodology:wiener_filter}
Consider a NN in which each node of the input layer is directly and linearly connected to each node of the output layer, as shown in \cref{fig:ZHL}. When minimising the MSE loss in \cref{eq:MSE_loss_via_posterior} for this linear NN, the resulting estimator is given by the WF,
\begin{equation}
    \est_i^\wf(\inp) = \sum_{j} \, w^\wf_{ij} \, \inp_j + b^\wf_i \,,
    \label{eq:NN_WF_estimator}
\end{equation}
with the weight matrix and bias vector
\begin{align}
    \mw^\wf &= \mC_{\targ\inp} \, \mC_{\inp\inp}^{-1} \,,
    \label{eq:WF_weights} \\
    \vb^\wf &= \langle \vtarg \rangle - \mw^\wf \, \langle \vinp \rangle \,.
    \label{eq:WF_biases}
\end{align}
Here, $\mC_{\targ\inp}$ is the cross-covariance matrix between target and input fields, $\mC_{\inp\inp}$ is the auto-covariance matrix of the input fields, and $\langle \vtarg \rangle$ and $\langle \vinp \rangle$ are the unconditional means of the training set target and input fields, respectively,
\begin{alignat}{2}
    C_{X_i Y_j} &= \langle X_i Y_j \rangle - \langle X_i \rangle \langle Y_j \rangle& \!\!\!\!\!\!\forall \vect{X},\vect{Y} &\in \{\vinp,\vtarg\} \,,
    \label{eq:field_covariance} \\
    \langle X_i \rangle = &\,\frac{1}{M} \, \sum_{\alpha=1}^M \, X^\alpha_i \xrightarrow{M \rightarrow \infty} \sum_\vect{X} \, P(\vect{X}) \, X_i& \forall \vect{X} &\in \{\vinp,\vtarg\} \,,
    \label{eq:mean_field} \\
    \langle \targ_i \inp_j \rangle = &\frac{1}{M} \, \sum_{\alpha=1}^M \, \targ^\alpha_i \inp^\alpha_j \xrightarrow{M \rightarrow \infty} \sum_{\vinp,\vtarg} \, P(\vinp,\vtarg) \, \targ_i \inp_j
    \label{eq:target_input_correlation} \\
    \langle \inp_i \inp_j \rangle = &\frac{1}{M} \, \sum_{\alpha=1}^M \, \inp^\alpha_i \inp^\alpha_j \xrightarrow{M \rightarrow \infty} \sum_\vinp \, P(\vinp) \, \inp_i \inp_j
    \label{eq:input_input_correlation} \,.
\end{alignat}
Only if the prior and likelihood are Gaussian, the WF agrees with the mean of the posterior. Otherwise the mean of the posterior is nonlinear in the input $\vinp$.

Note that, despite the mathematical equivalence shown above, all WF reconstructions shown in \cref{sec:results} are computed via the conventional implementations of the WF described in \cref{sec:WF_computation}, to avoid any potential numerical inaccuracies in our reference WF reconstructions.

\subsection{Linear regressions}
\label{sec:methodology:estimator_properties}
Although for non-Gaussian data, the mean posterior and WF estimators are not equivalent, the linear regression of the target fields on either of these two estimates,
\begin{equation}
    \targ^\lin_i(\est_i) = \beta \, \est_i + \gamma + \varepsilon_i \,,
\end{equation}
yields a slope $\beta = 1$ and an offset $\gamma = 0$, plus a random scatter $\varepsilon_i$. To show this, we start with the general expressions for the slope and offset obtained by least squares optimization,
\begin{align}
    \beta &= \frac{\frac{1}{N} \sum_{i=1}^N \langle\targ_i\est_i\rangle - \frac{1}{N^2} \sum_{i,j=1}^N \langle\targ_i\rangle\langle\est_j\rangle}{\frac{1}{N} \sum_{i=1}^N \langle\est_i^2\rangle - \frac{1}{N^2} \bigl(\sum_{i=1}^N \langle\est_i\rangle\bigr)^2} \,,
    \label{eq:linear_regression_slope} \\
    \gamma &= \frac{1}{N} \sum_{i=1}^N \bigl(\langle \targ_i \rangle - \beta \, \langle \est_i \rangle\bigr) \,.
    \label{eq:linear_regression_offset}
\end{align}
Both estimators preserve the unconditional mean of the target fields,
\begin{align}
    \bigl\langle \vest^\mse \bigr\rangle &= \sum_\vinp P(\vinp) \, \vest^\mse(\vinp) = \sum_{\vtarg,\vinp} P(\vinp) \, P(\vtarg|\vinp) \, \vtarg = \langle \vtarg \rangle \,,
    \label{eq:unconditional_mean_mean_posterior} \\
    \bigl\langle \vest^\wf \bigr\rangle &= \sum_\vinp P(\vinp) \, \vest^\wf(\vinp) = \sum_\inp P(\vinp) \, \bigl(\mw^\wf \vinp + \vb^\wf\bigr) = \langle \vtarg \rangle \,,
    \label{eq:unconditional_mean_WF}
\end{align}
following from \cref{eq:mean_posterior_estimator,eq:NN_WF_estimator,eq:WF_biases}. Furthermore,  the two correlation functions in the enumerator and denominator of \cref{eq:linear_regression_slope} are equal, as can be see as follows,
\begin{align}
    \bigl\langle\bigl(\est^\mse_i\bigr)^2\bigr\rangle &= \sum_\vinp \, P(\vinp) \, \est^\mse_i(\vinp) \, \est^\mse_i(\vinp) \\
    &= \sum_{\vinp,\vtarg} \, P(\vinp) \, P(\vtarg|\vinp) \, \targ_i \, \est^\mse_i(\vinp) \\
    &= \sum_{\vinp,\vtarg} \, P(\vinp,\vtarg) \, \targ_i \, \est^\mse_i(\vinp) \\
    &= \bigl\langle\targ_i\est^\mse_i\bigr\rangle \,, \\
    \bigl\langle\bigl(\est^\wf_i\bigr)^2\bigr\rangle &= \sum_{j,l=1}^N w^\wf_{ij} w^\wf_{il} C_{\inp_j\inp_l} + \langle \targ_i \rangle^2 \\
    &= \sum_{l=1}^N w^\wf_{il} C_{\targ_j\inp_l} + \langle \targ_i \rangle \langle \est_i \rangle \\
    &= \bigl\langle\targ_i\est^\wf_i\bigr\rangle \,.
\end{align}
Here, we have used the estimator definitions in \cref{eq:mean_posterior_estimator,eq:NN_WF_estimator,eq:WF_weights,eq:WF_biases} together with \cref{eq:field_covariance,eq:unconditional_mean_WF}.
By inserting the two relations $\langle\est_i\rangle = \langle\targ_i\rangle$ and $\langle\est_i^2\rangle = \langle\targ_i\est_i\rangle$ into \cref{eq:linear_regression_slope,eq:linear_regression_offset}, we then immediately infer a slope $\beta = 1$ and offset $\gamma = 0$. In other words, for both the mean posterior and the WF estimators, the linear regression of the target fields on the estimated fields recovers the latter up to random scatter, $\targ^\lin_i(\est) = \est_i + \varepsilon_i$. The variance of the scatter, however, is smaller for the MSE-minimising mean posterior estimate (unless the input and target fields are Gaussian).

The inverse linear regression of the estimate on the target does \emph{not} recover the target plus a random scatter, $\est^\lin_i(\targ) \neq \targ_i + \varepsilon_i$. This is because the correlation functions appearing in the slope of the inverse regression are not equal, $\langle\targ_i^2\rangle \neq \langle\est_i\targ_i\rangle$. 

A slope of unity for the linear regression of $\targ_i$ on $\est^\mse_i$ is a direct consequence of the mean posterior estimate being unbiased in the sense that $\langle\targ_i|\est^\mse_i\rangle = \est^\mse_i$, i.e.~the average target value conditional on the estimated value equals the latter. This follows from the definition of $\est^\mse_i$ given in \cref{eq:mean_posterior_estimator}. For non-Gaussian fields, the WF does \emph{not} in general share this property, $\langle \targ_i|\est^\wf_i \rangle \neq \est^\wf_i$, despite a slope of unity of $\targ_i$ on $\est^\wf_i$.

\begin{figure*}
\centering
    \includegraphics[width=\textwidth]{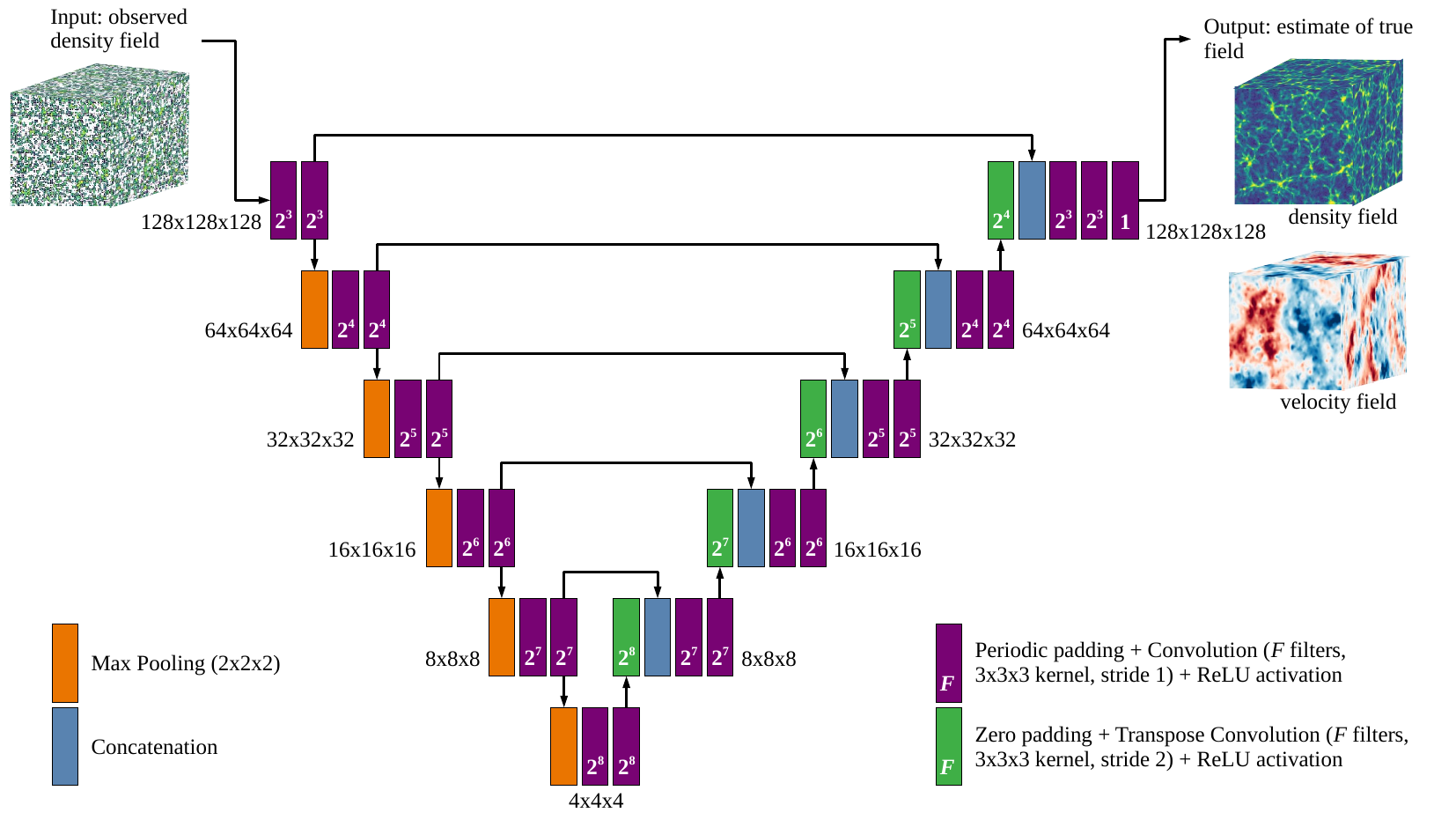}
    \caption{Schematic of the \Ae\  U-Net architecture used in this paper for reconstructing the  
    underlying 3D density and velocity field from a noisy density field.  See \cref{sec:methodology:autoencoder} for further details.}
    \label{fig:Unet_architecture}
\end{figure*}

\subsection{Autoencoder}
\label{sec:methodology:autoencoder}
In a densely connected NN every node in a given layer is connected to all nodes in the previous layer, i.e, indices the $i$ and $j$ in \cref{eq:neural_net_node_value} run over all nodes in each layer.
Dense NNs are, in principle, capable of learning
the complete hierarchy of correlations involved in the 
determination of the weights $w$ and biases $b$ from the training set.
Unfortunately, dense NNs do not scale up efficiently  for large data. For example, reconstructing 3D cosmological fields on a $128^3$ grid becomes computationally prohibitive, as it requires repeated numerical manipulations of huge weight matrices of at least $128^3\times 128^3$ elements. Therefore, we use a convolutional NN, instead, which only connects nearby nodes between adjacent layers, thus substantially reducing the number of weights.
Increasing the depth of a convolutional NN, i.e.~increasing the number of hidden layers, ensures that it nevertheless captures long-range correlations in the training data. This is an essential feature for  the reconstruction of large-scale structures. Furthermore, as per the \emph{universal approximation theorem}, a feedforward NN is capable of approximating any function to any desired degree of accuracy, provided there are enough hidden nodes \citep{hornik_multilayer_1989}. Increasing the depth thus also improves the ability of the NN to accurately express the nonlinear mapping from the input to the mean posterior estimate of the target.

For our purposes we adopt an autoencoder (\Ae) convolutional NN \citep{kramer_nonlinear_1991}, which consists of two parts: First an \emph{encoder} compresses the main features of an input field into a reduced latent space representation. Then a \emph{decoder} expands this compressed information into the estimate of the target field.\footnote{An \Ae\ with  linear activations and an MSE loss function is  equivalent to Principal Component Analysis (PCA) \citep{goodfellow_deep_2016}.} AEs have already proven effective for image denoising and inpainting \citep[e.g.][]{xie_image_2012}. 
 
\Cref{fig:Unet_architecture} visualizes our specific choice of the \Ae\ architecture, which follows a U-net structure \citep{ronneberger_u-net_2015, inpaintingAyush2020}.
The encoder part of this architecture compresses a $128 \times 128 \times 128$ observed density field to a latent space representation of $4\times 4 \times 4$ per filter in a total of 256 filters. The decoder part then expands the compressed data to a $128 \times 128 \times 128$ estimate of the underlying density or velocity field. Both parts are performed over several steps. Each encoding step consists of two 3D convolutions followed by a 3D max pooling layer. The convolutions are performed with periodic padding instead of the default zero padding, as the training data boxes are generated with periodic boundary conditions. The chosen numbers of convolutional filters per layer are specified in the figure. Each convolution is followed by a rectified linear unit (ReLU) activation, which introduces nonlinearity by setting all negative node values to zero. The max pooling chooses the maximum value of each $2\times 2\times 2$ sub-grid and hence halves the data size per spatial dimension. Each decoding step consists of two periodically padded 3D convolutions followed by a zero-padded 3D transpose convolution, all using ReLU activations. The transpose convolution simultaneously doubles the data size per spatial dimension. The encoder and decoder are linked via concatenations called skip connections, as indicated in the figure. These skip connections between partially encoded and decoded data help to preserve the spatial localisation of features in the data.

The number of trainable weights and biases in our \Ae\ is $\sim 6.6\times 10^6$, and thus significantly smaller than the total number of data points used in the training: $\sim 1.0\times10^9$ for density reconstructions and $\sim 2.1 \times 10^9$ for velocity reconstructions. We implement the \Ae\ in Keras \citep{keraschollet2015} using TensorFlow \citep{tensorflow2015-whitepaper} as the backend. For each of the three considered types of fields (cf.~\cref{sec:data}) the training took less than two hours on a single NVIDIA A100 GPU with 40 GB of memory.

\begin{table*}
    \centering
    \caption{Summary of  the training, validation and test data properties for the different types of fields.}
    \label{tab:training_data}
	\begin{tabular}{ @{}lcccccc@{}} 
        \toprule
		 & \textbf{Box length} & \textbf{Grid cells} & & \textbf{Samples} & & \textbf{Mini-batch size} \\
		\cmidrule{4-6}
		 & $[\hmpc]$ & & Training & Validation & Test & \\
        \midrule
        \textbf{Gaussian} & 300 & $128^3$ & 500 & 50 & 20 & 5 \\
    	\textbf{2LPT} $\delta$ & 300 & $128^3$ & 500 & 50 & 20 & 5 \\
    	\textbf{2LPT} $v_z$ & 300 & $128^3$ & 1000 & 10 & 20 & 5 \\
    	\bottomrule
    \end{tabular}
\end{table*}

\section{Data}
\label{sec:data}
The main ingredient for a successful application of any NN is the availability of suitable training data samples. Here we employ 2LPT fields to approximate the dynamics of particles in a cosmological background. In addition, we demonstrate the equivalence of the \Ae\ and the WF for Gaussian fields.

For each type of field, we divide the entire data into training, validation, and test sets, as given in \cref{tab:training_data}. 
The training set is used to learn the \Ae\ weights and biases by minimizing the MSE loss function during the training process, as described in \cref{sec:methodology}. It is split into into mini-batches of five samples each. Increasing the mini-batch size reduces fluctuations in the loss function, but is limited by the GPU and CPU memory sizes. The validation set is used to independently evaluate the MSE loss at the end of each training epoch. 
After the MSE loss converged, the test set is used to compare the statistics of the reconstructed fields, e.g.~the PDF and power spectrum.
As we shall see, in all cases convergence  is largely reached  by epochs 30\,--\,40. The specific epochs at which the reconstructions of the various fields are performed differ in each case. 
For comparison, we compute the WF reconstructions, as explained in \cref{sec:WF_computation}, for the same test set.

\subsection{Gaussian fields}
\label{sec:data:gaussian}

We have generated an ensemble of underlying random Gaussian density fields
in a $128^3$  cubic grid with side length 
$300\,\hmpc $ and periodic boundary conditions (cf.~\cref{tab:training_data}). 
These fields are generated using the 
power spectrum of the 2LPT underlying density fields (see below), to facilitate 
 the comparison between both types of fields. 
The underlying fields have an rms value of $1.7$. The observed fields are obtained by adding uncorrelated random Gaussian noise with mean  $\mu=0$ and  rms $\sigma=4.2$ (see \cref{tab:density_statistics}). The Gaussian noise rms was chosen to match the Poisson noise rms of the 2LPT fields described below.

\subsection{Nonlinear fields: 2LPT}
\label{sec:data:2LPT}

Ideally, we would like  to test the \Ae\ 
reconstruction on realistic cosmological simulations. 
Here we are only interested in structures on large scales, however, where strongly nonlinear effects such as shell crossing or virialization are not relevant.
Thus, it suffices  to resort to  2LPT, 
which provides a  reasonable approximation to the nonlinear evolution of cosmological fields up to the stage of shell crossing. This perturbation theory can be applied easily to  generate sufficiently many training and validation samples (cf.~\cref{tab:training_data}).

In our application, particles are initially placed on the vertices of a $128^3$ cubic grid with 
comoving side length $300 \, \hmpc$. The initial conditions to be fed to the 2LPT relations are Gaussian with a linear power spectrum of the Planck-18 $\Lambda$CDM cosmology \citep{aghanim_planck_2020-1}. The 2LPT solution then provides the velocity, $\vvel$, and (evolved) Eulerian coordinates, $\vr$, of the particles as a function of the (initial) Lagrangian coordinates, $\vq$. Further details on the generation of the 2LPT fields are given in \cref{sec:2LPT_data_generation}.

Given its Eulerian coordinate, $\vr$,
each  particle is assigned to its nearest grid cell. The number of particles in a cell is then used to obtain the density in real space in that cell. 
For the underlying true density we 
use all particles in the box, while for the observed field (without RSD) we randomly select a subset of the particles matching the desired mean galaxy number density $\bar{n}$. Unless specified otherwise, we use $\bar{n} = 5 \times 10^{-3} \, \hmpcdens$. This mean galaxy density and the chosen box size are roughly comparable to the characteristics of the 2MASS Redshift Survey \citep{huchra_2mass_2012}.

The observed density including RSD  is obtained similarly but using the 
Eulerian redshift-space coordinate $\vs$ instead of $\vr$, computed as
\begin{equation}
    \vs=\vr + \frac{v_z}{H} \, \vhz \,.
\end{equation}
Here, $v_z$ is the  peculiar velocity component in the $z$-axis,
chosen to represent  the line-of-sight direction, and $H$ is the Hubble constant.

We also compute the velocity field as a function of the Eulerian coordinates by averaging the particle velocities at each (Eulerian) grid cell.

In the analysis below we assess the ability of the \Ae\ at reconstructing the underlying (Eulerian) density as well as Lagrangian and Eulerian velocity fields from the observed density fields either with or without RSD.

\section{Results}
\label{sec:results}

\begin{figure*}
    \includegraphics[width=\textwidth]{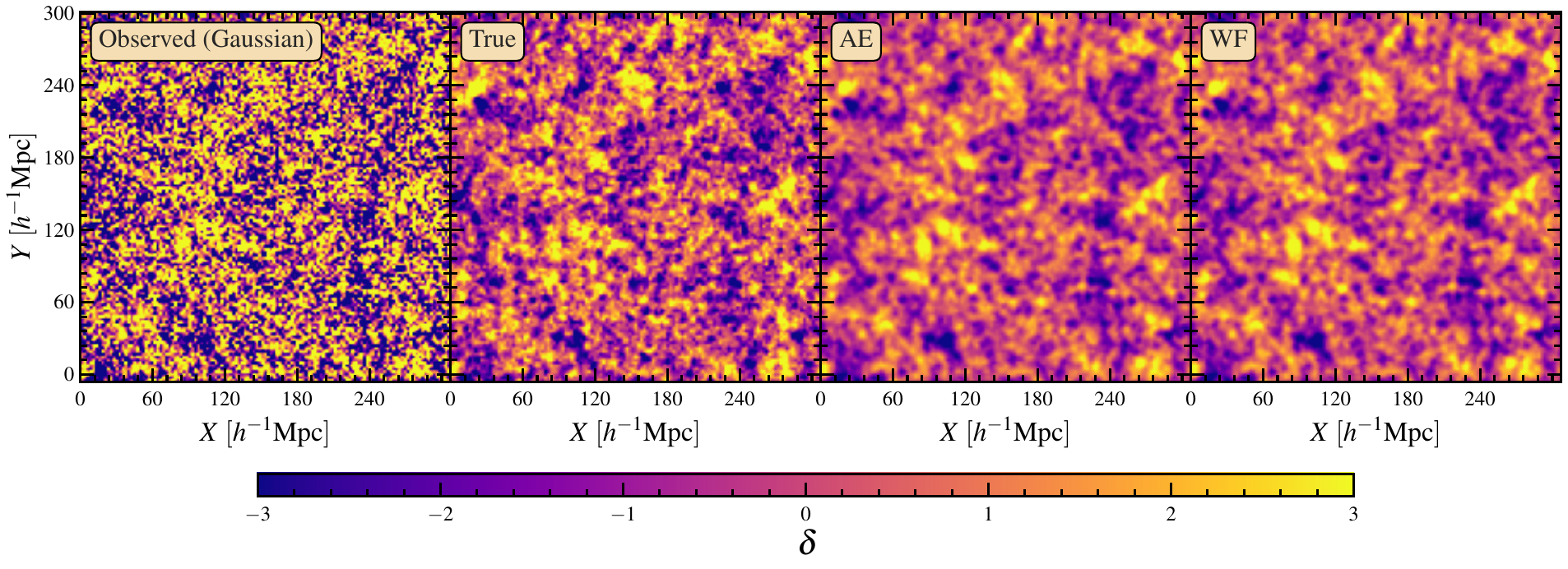} %
    \caption{ Comparisons of the reconstructions from noisy Gaussian fields using the \Ae\ and standard WF. Shown, as indicated in the figure, are the observed, true, \Ae\ and WF reconstructions for one of the samples from the test set.  The \Ae\ and WF reconstructions in this Gaussian case are visually indistinguishable.
    }
    \label{fig:gaussian_fields}
\end{figure*}

\begin{figure*}
  	\centering
  	\begin{tabular}{rr}
    \imagetop{\includegraphics[width=0.98\columnwidth]{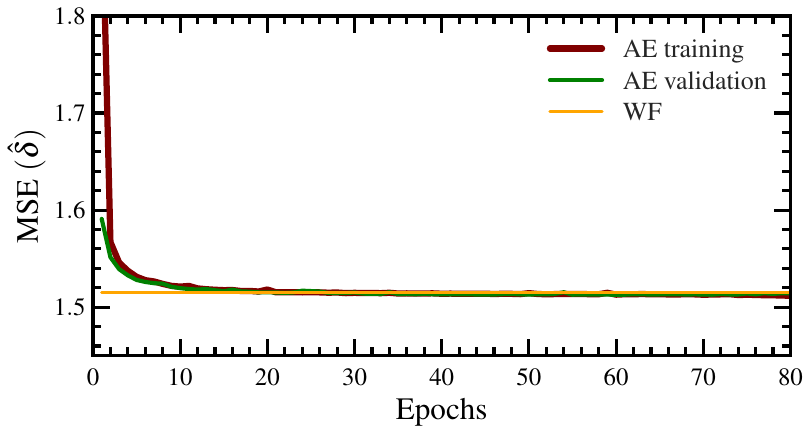}} & 
  	\imagetop{\includegraphics[width=0.95\columnwidth]{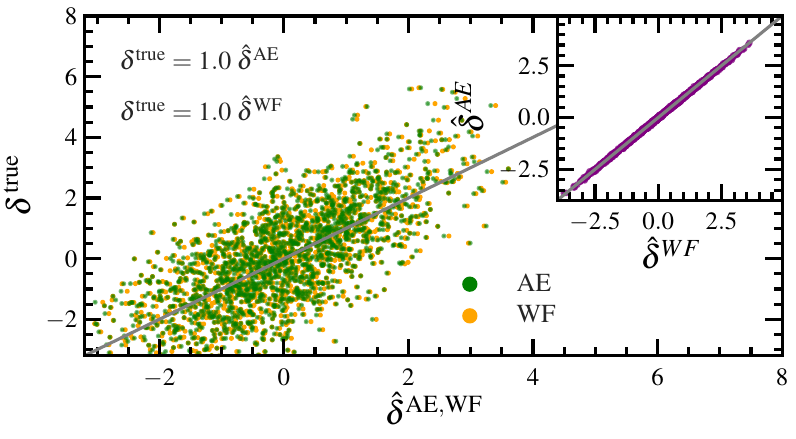} \hspace*{0.01\columnwidth}} \\
  	\imagetop{\includegraphics[width=0.98\columnwidth]{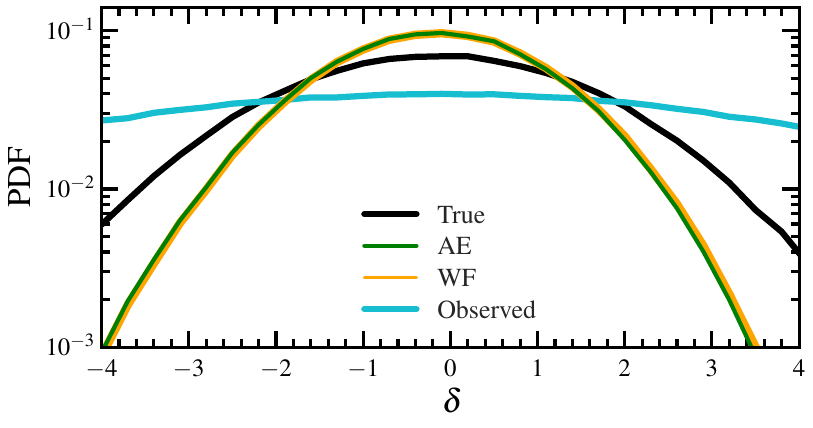} \hspace*{0.001\columnwidth}} & 
 	\imagetop{\includegraphics[width=0.98\columnwidth]{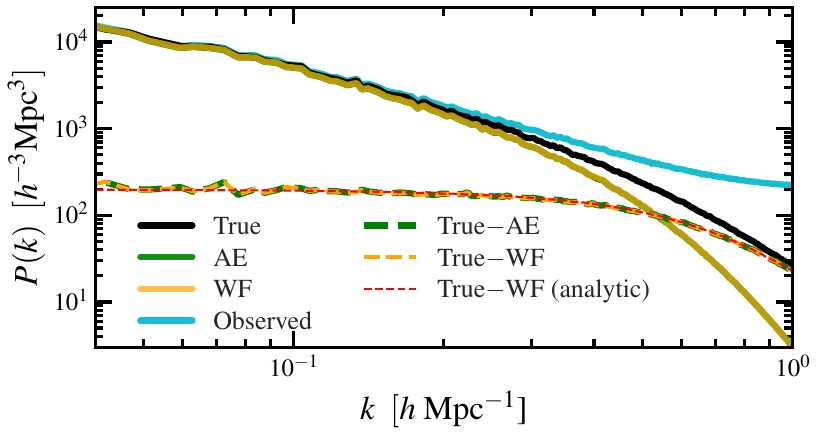}}
  	\end{tabular}
	\caption{Various statistics of the true and reconstructed Gaussian fields.  {\textit{Top left:}} the  MSE loss function (\cref{eq:MSE_loss}) versus the  training epoch. The MSE value of the WF reconstruction is plotted as the horizontal amber line for comparison. The nonlinear AE converges to the WF MSE for the case of the Gaussian fields. \textit{Top right:}  a point-by-point scatter plot of true versus reconstructed densities at a random selection of grid points for the fields shown in the previous figure. The inset plot demonstrates the near-perfect equivalence between the  nonlinear \Ae\ and Wiener fields. \textit{Bottom left:}  the PDFs of the density computed from the test set, as indicated in the panel. \textit{Bottom right:} 
	average power spectra of the density fields as well as the power spectra of {the} residual between the true and the reconstructed {fields}.
    }
	\label{fig:gaussian_statistics}
\end{figure*}

\subsection{Reconstruction of a Gaussian field}
\label{sec:results:gaussian}
As we have seen in \cref{sec:methodology:MSE}, in an ideal case of Gaussian underlying and observed fields, the 
WF estimate  of the underlying field coincides with both the  mean field and the median field given the observations. All three fields  are identical and yield  the minimum variance solution for the given set of samples. 
Since the nonlinear AE aims at retrieving the mean field, it should thus coincide with the WF estimate in the fully Gaussian case. 

In this subsection, we test how well the \Ae\ converges to the WF field for the Gaussian case, 
where the observed field is provided in a cubic box.
The WF estimate corresponding to each observed field is obtained  in Fourier space according to the expression in \cref{eq:WF_density_unmasked}.

\Cref{fig:gaussian_fields} displays, as heatmaps,  observed, true and reconstructed density fields for one of the realizations 
in a slice through the cubic box.
The effect of the noise in the observed field  is   pronounced in comparison to the true underlying field.
The WF reconstruction is visually indistinguishable from the full nonlinear \Ae\ field, implying that the \Ae\ is indeed 
capturing the mean field, as it should. The reconstructed field appears blurred and does not  capture the details of the small-scale structure of the true field. This is in accordance with 
the small-scale suppression (large wavenumbers) implied by the WF expression \cref{eq:WF_density_unmasked}. 
  
For an estimate $\hat \delta$ of the true density $\delta$, we compute the  MSE
defined in accordance with \cref{eq:MSE_loss},
\begin{equation}
   \textrm{ MSE}(\hat \delta) =\frac{1}{M N}\sum_{\alpha=1}^M \sum_{j=1}^N \left(\hat \delta^\alpha_j -\delta^\alpha_j\right)^2 \,,
\end{equation} 
where $\alpha=1,\dotsc, M$ indicates the sample field in the training (or validation) set, and  $j=1,\dotsc, N $ is the grid cell index. 

\Cref{fig:gaussian_statistics} provides  a quantitative assessment of the \Ae\ and WF reconstruction and their mutual agreement.  The top left panel  plots curves of 
the MSE versus training epoch for the \Ae\ reconstruction for 
the training  ({brown}) and validation (green) sets. 

The horizontal amber line indicates the MSE value computed from the WF reconstructed density.
The  MSEs of the \Ae\  for both training and validation sets start declining rapidly until they saturate at the WF value at epoch $\sim 20$. Initially, the validation MSE is lower than the training MSE because the former is computed at the end of a training epoch, whereas the latter is a running average over all mini-batches during the epoch.
Fluctuations in the MSE continue to appear but  at an insignificant level. The main reason for these fluctuations is the finite number of samples  used in the mini-batch training scheme as discussed in \cref{sec:data}.

The top right panel in the same figure is a point-by- point comparison of the true versus reconstructed densities from the \Ae\ (green points) and WF (amber).
For clarity, densities on a random subset of grid cells are plotted.
The slope of the linear  regression of the true on the reconstructed values is unity, in agreement with the discussion in \cref{sec:methodology:estimator_properties}.
The \Ae\ and WF points are almost identical, as confirmed by the inset showing the \Ae\ versus WF densities. 

The  PDFs are shown in the bottom left panel. 
The PDF of the reconstructed densities is symmetric with respect to $\delta=0$ but they are  significantly different from true Gaussian PDF due to the suppression of signals at the tails of the distribution. The \Ae\ and WF PDFs agree remarkably well, in accordance with the scatter plot in the top right panel.  

\Cref{fig:gaussian_statistics} also shows, in the bottom right panel, 
power spectra, $P(k)$, of the various density fields.
The curves show the mean spectra from  samples in the test set. 
At small wavenumbers (large scales), $k\lesssim 0.15 \, \hmpcinv$, both the \Ae\ (green curve)  and the WF (amber) 
power spectra match the underlying true $P(k)$ (black) extremely well. 
At larger wavenumbers, the two reconstructed spectra are indistinguishable in the plot and are suppressed compared to the true $P(k)$.
Due to the presence of noise, the observed $P(k)$ (blue) lies markedly above all the other curves on small scales.  
The dashed curves are the   power spectra computed on the residual fields, $\delta-\hat \delta$.
They clearly match the  analytic expectation  (dashed red) for the residuals following from \cref{eq:WF_density_unmasked}.

For easy comparison, the values of various statistical properties (variance, MSE, skewness and regression slopes) of the true, observed and reconstructed density fields are listed in \cref{tab:density_statistics}.

\begin{figure*}
    \includegraphics[width=\textwidth]{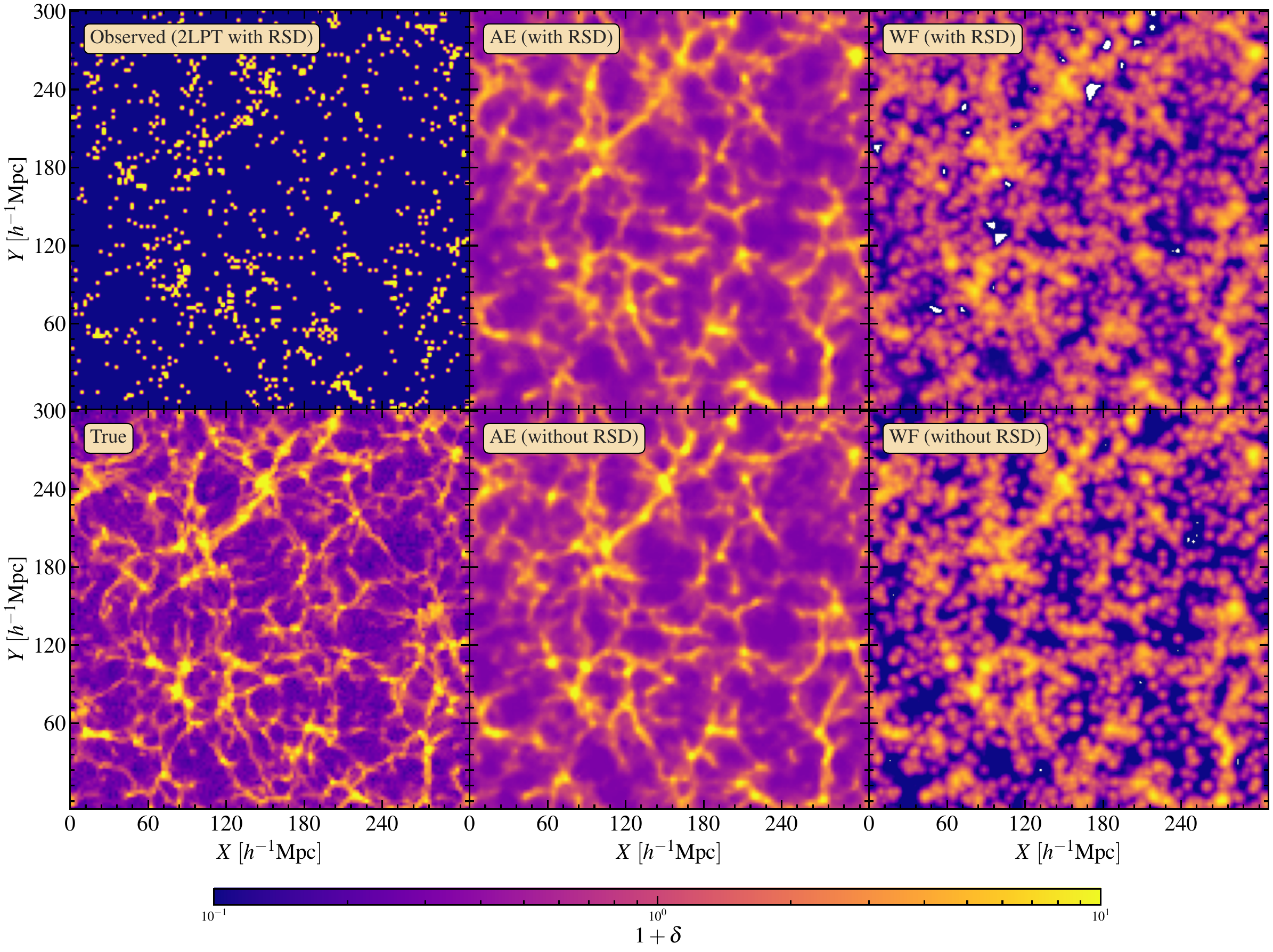}
    \caption{Reconstructions of  density fields in a slice through one of the test set {realizations} generated using 2LPT. \textit{Top row:} the observed density contrast (left panel)  and \Ae\ and WF reconstructions including RSD, as indicated in the figure. \textit{Bottom  row:} the true density (left panel) and the reconstructed fields  in real space.}
    \label{fig:2LPT_density_fields}
\end{figure*}

\begin{figure*}
  	\centering
  	\begin{tabular}{rr}
  	\imagetop{\includegraphics[width=0.98\columnwidth]{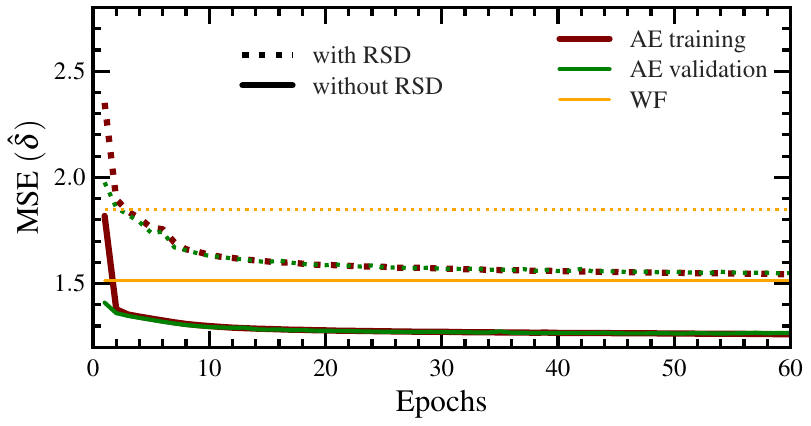}} &
  	\imagetop{\includegraphics[width=0.97\columnwidth]{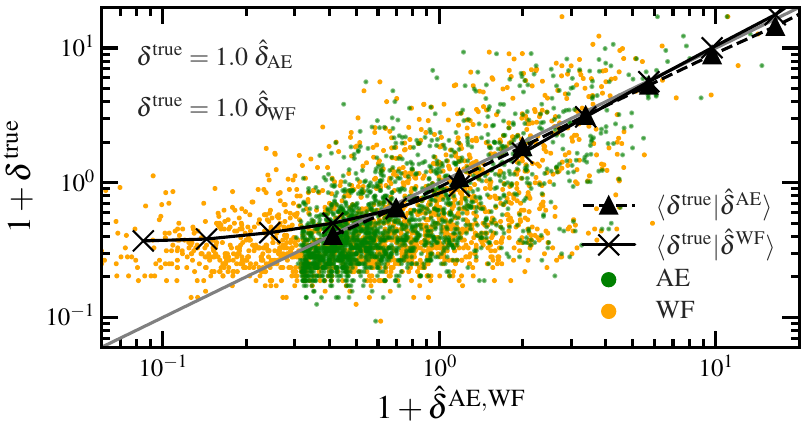} \hspace*{0.012\columnwidth}} \\
  	\imagetop{\includegraphics[width=0.975\columnwidth]{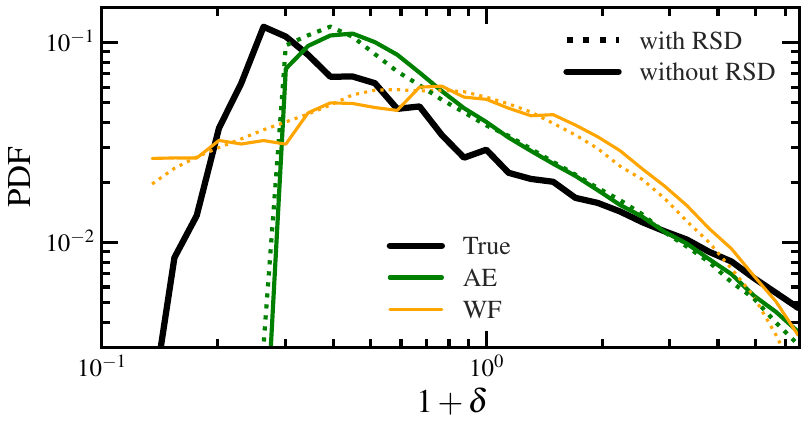} \hspace*{0.008\columnwidth}} &
    \imagetop{\includegraphics[width=0.98\columnwidth]{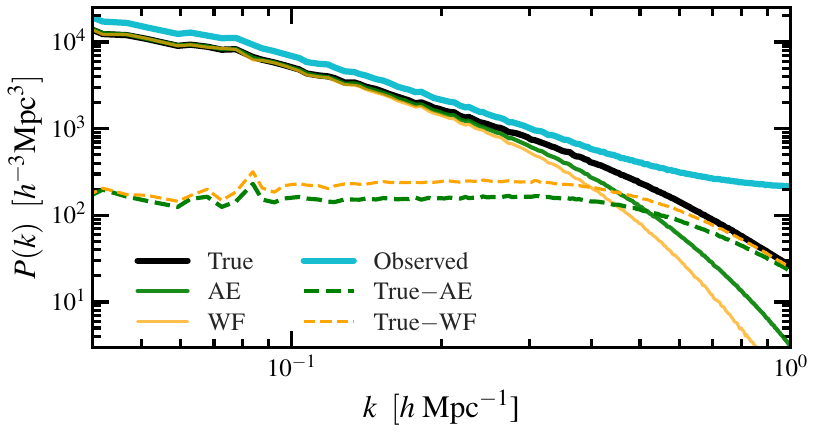}}
  	\end{tabular}
      \vskip -0.02cm
	\caption{Statistics of the \Ae{-} and WF{-}reconstructed density fields for the  2LPT  data. 
    {\textit{Top left:}} the MSE{s} of training (red) and validation (green) sets versus {the} epoch. MSEs for the  WF (amber) reconstruction are  shown as horizontal lines. 
    Solid curves  refer to reconstructions from {the} observed density  {without} RSD, while dotted curves correspond to {the case with} RSD.
	\textit{Top right:} scatter plot of  true versus reconstructed densities (with RSD), as indicated in the figure, for a random subset of grid cells in {the test set realization shown in \cref{fig:2LPT_density_fields}}.
	{The slopes} of linear regressions are indicated in the figure. Triangles and crosses are the mean $\delta^\textrm{true}$ in a given bin of $\hat{\delta}^\textrm{AE}$ and $\hat{\delta}^\wf$, respectively, over all grid cells. 
	\textit{Bottom left:} PDFs, as indicated in the figure. 
	\textit{Bottom right:} mean of power spectra  computed from the true{, reconstructed and observed} density fields {with RSD} in the test samples. {The} power spectra of residual fields are also plotted.}
	\label{fig:2LPT_density_statistics}
\end{figure*}

\begin{table*}
    \centering
    \caption{Statistical properties of the true, observed and reconstructed density fields, computed from the test set. 
    }
    \label{tab:density_statistics}
	\begin{tabular}{ @{}llccccc@{}} 
        \toprule
        \\
		\textbf{}  & \textbf{Field} & \textbf{Variance} & \textbf{MSE } & \textbf{Skewness}  & \textbf{Regression slope} &  \textbf{Regression slope} \\ 
				  & $\delta_{\mathrm {field}}$ & $\langle \delta_{\mathrm{field}} ^{2}\rangle $&  $ \langle (\delta_{\mathrm {true}}-\delta_{\mathrm{field  }})^{2} \rangle$ & ${\langle \delta_{\mathrm{field  }}^{3}\rangle}/{\langle \delta_{\mathrm{field  }} ^{2}\rangle^{3/2} } $ & \textbf{ $\delta_{\mathrm {true}}$} {vs} {$\delta_{\mathrm {field}}$} &  \textbf{$\delta_{\mathrm {field}}$} {vs} {$\delta_{\mathrm {true}}$ } \\ \\
       \toprule
        \multirow{4}{*}{\textbf{Gaussian }} 
	    & True & $ 3.0$ &  $-$  & $0 $ & $-$ & $- $ \\
	     & Observed & $18$ & $  15$ & $ 0$ & $0.17$ & $1.0$\\
    		&Autoencoder & $ 1.5 $ & $ 1.5 $ & $0$ &$1.0$ & $0.50$\\ 
    		& Wiener filter & $ 1.5 $ & $ 1.5$ & $0$ & $ 1.0$ & $0.50$\\ 
    		\midrule
    		\multirow{4}{*}{\textbf{2LPT without RSD }} 
	    &True & $ 3.1 $ & $ -$ & $6.1 $ & $-$ & $-$\\
	    & Observed & $ 18$  &  $15$ & $5.2 $& $0.17$ & $1.0$\\
	&Autoencoder & $ 1.7$ & $1.3 $ & $5.8 $ & $1.0$ & $0.57$\\ 
    		& Wiener filter & $1.5$&$1.5$ & $3.5$ & $1.0$ & $0.50$ \\
     		\midrule
    	\multirow{3}{*}{\textbf{2LPT with RSD} } & Observed & $18$ & $17$ & 5.0 & $0.12$ & $0.71$ \\
    		&Autoencoder & $ 1.6 $ & $1.6 $ & $ 4.9 $  & $ 0.99$ & $ 0.50 $  \\
    		& Wiener filter & $1.2$ & $1.9 $ & $3.2$ & $0.99$ & $0.39 $\\
    		\bottomrule
    \end{tabular}
\end{table*}

\subsection{2LPT density reconstruction}

\label{sec:results:2LPT_density}
After demonstrating the equivalence of the \Ae\ with the WF for Gaussian field reconstructions, we now explore the differences between the two estimators for the non-Gaussian 2LPT fields, to verify that the \Ae\ reconstruction displays the properties expected from the nonlinear mean posterior estimator. In this subsection, we consider the reconstruction of the underlying density field from the noisy observed density.

In \cref{fig:2LPT_density_fields} we show a slice through the $X-Y$ plane for one randomly chosen density realization {from the test set}. As indicated in the figure, the individual panels display the true and observed fields as well as the \Ae\ and WF reconstructions. Due to the visual similarity between the observed density fields with and without RSD, we only show the one with RSD for reference. The reconstructions are shown for both cases, however. The true 2LPT density exhibits the familiar filamentary structure of the cosmic web. Despite having the same power spectrum as the Gaussian density shown in \cref{fig:gaussian_fields}, the visual difference between the two fields is striking. The observed 2LPT density represents the galaxy counts following from Poisson sampling. For the chosen mean galaxy density of $\bar{n} = 5 \times 10^{-3} \, \hmpcdens$, most grid cells do not contain any galaxies, resulting in the sparse observed density field. 

The \Ae\ and WF reconstructions  differ markedly in their visual characteristics. The \Ae\ reconstruction appears as a blurry  version of the true field; it recovers all the major filaments and preserves the intricate connectivity of the underlying web. In contrast, the WF seems  like a smoothed version of the observed field, i.e.~a superposition of individual smoothing kernels for each observed galaxy. Although the WF still recovers the large-scale structures of the true density, the connectivity of the underlying web is 
not as highlighted as in the \Ae. Furthermore, the WF can yield unphysical negative densities in voids, since it is not designed to enforce positivity. In contrast, the \Ae\ approximates the actual mean of the posterior distribution, given in \cref{eq:mean_posterior_estimator}, and hence never yields negative densities. In addition, this is numerically guaranteed by the ReLU activation used in the final layer of the \Ae, as described in \cref{sec:methodology:autoencoder}. Instead, the \Ae\  slightly overestimates the densities in voids, as it smooths over the small-scale structures. For both \Ae\ and WF, the reconstructions from the observed density with and without RSD are visually similar. With RSD, however, the highest density peaks are generally underestimated to a larger extent. The volume in which the WF predicts negative densities  also increases for the case with RSD.

In \cref{fig:2LPT_density_statistics} we present a number of quantitative comparisons between the \Ae\ and WF reconstructions. The top left panel shows the \Ae\ MSE for the training and validation data sets versus the training epoch. For comparison, we plot the WF MSE as a horizontal line. The \Ae\ MSE quickly drops below the WF MSE within the first few epochs, and then converges to its lowest value after approximately 40\,--\,60 epochs. 
Apart from the very few first epochs, the MSEs for the training and validation sets agree, thus showing no signs of overfitting. The MSE for the reconstruction with RSD is larger than without RSD, about as much as the difference between the MSEs of WF and \Ae. The precise numerical MSE values can be found in \cref{tab:density_statistics}.

The top right panel https://www.overleaf.com/project/621e05c6c0f594dde610d2d0of \cref{fig:2LPT_density_statistics} is a scatter plot of the true density versus the \Ae\ and WF reconstructions for a random subset of grid cells in {the realization with RSD shown in \cref{fig:2LPT_density_fields}}
 (the scatter plot without RSD looks similar). The most notable difference is that the \Ae-reconstructed density never drops below $\sim 0.3$, whereas the WF-reconstructed density reaches values far below $0.1$ (in fact below zero, which is just not visible on the logarithmic axis), in agreement with the visual analysis of the fields in \cref{fig:2LPT_density_fields}.

At higher densities, the scatter of the \Ae\ points is slightly tighter than {in the} WF reconstruction; otherwise they are similar. Despite the  differences in the low-density scatter, the linear regression slope is found to be very close to unity for both the \Ae\ and the WF reconstructions. As confirmed by the average regression slopes given in \cref{tab:density_statistics}, this is a general property and not only true for the one realization shown here. In the same table we also provide the average slopes of the inverse regression, reconstructed versus true density, which yields values smaller than unity for both \Ae\ and WF.
{We also plot} the average true density in bins of reconstructed  densities, i.e.~the conditional average of true given reconstructed densities.
The connected  triangles representing $\langle \delta^\textrm{true}|\delta^\textrm{AE } \rangle$  lie very close to the diagonal grey line. Despite  {the} slope of unity for the linear regression of true {versus} WF density, the 
deviations {of the connected crosses, $\langle \delta^\textrm{true}|\delta^\wf \rangle$,} from a straight line are evident. 
These results are in agreement with the theoretical expectation discussed in \cref{sec:methodology:estimator_properties}.

In the bottom left panel of \cref{fig:2LPT_density_statistics} we plot the PDFs of the true and reconstructed densities, computed from the test set. The skewed true PDF (black solid) demonstrates the significant non-Gaussianity of the 2LPT density field. The PDF of the \Ae\ reconstruction (green) displays a good qualitative agreement, with the largest difference being the low-density cutoff at $\sim 0.3$ already seen in the field slices and the scatter plot. In contrast, the WF (amber) yields a more symmetric PDF, which extends to much lower densities than the true PDF, and peaks closer to the mean density than both the true and \Ae\ PDFs. This is confirmed by the lower skewness value of the WF-reconstructed density, given in \cref{tab:density_statistics}, compared to the very similar high skewness values found for the true and \Ae-reconstructed densities. The reconstructed PDFs are less sensitive to the presence of RSD than the MSE, as the difference between the reconstructed PDFs with or without RSDs is negligible. 

The bottom right panel in \cref{fig:2LPT_density_statistics} plots the power spectra of the true, reconstructed and observed density fields (solid lines), computed from the test set. As for the scatter plot, we only show the case with RSD. At small wavenumbers (large scales) $k \lesssim 0.15 \, \hmpcinv$, the true and reconstructed spectra closely agree, whereas the observed spectrum is boosted by the RSD \citep{kaiser_clustering_1987}. At larger wavenumbers both reconstructed spectra drop below the true $P(k)$, with the \Ae\ being closer to the true spectrum than the WF. At these wavenumbers, the observed power spectrum approaches a constant value corresponding to the noise level.

We additionally plot the power spectra of the residuals between the true and reconstructed density fields (dashed lines).
The residual spectra confirm the agreement with the \Ae\ and WF on small $k$. {On noise-dominated} scales (large $k$), they approach the true $P(k)$. On intermediate scales, the residual $P(k)$ of the \Ae\ falls below the WF.

Altogether, the visual inspection as well as the statistical properties of the \Ae-reconstructed density fields agree with the expected characteristics of the mean posterior estimator.

\subsubsection{Effect of masking}
\label{sec:results:2LPT_density:mask}
Galaxy surveys  cover {only} part of the sky, and even all-sky surveys suffer from obscuration by the disk of the Milky Way (the Zone of Avoidance). Studying the effect of partially masked observations on the reconstruction quality is thus important. We provide here a very basic test of the effect of a masked region in the observed density.

\begin{figure*}
    \includegraphics[width=\textwidth]{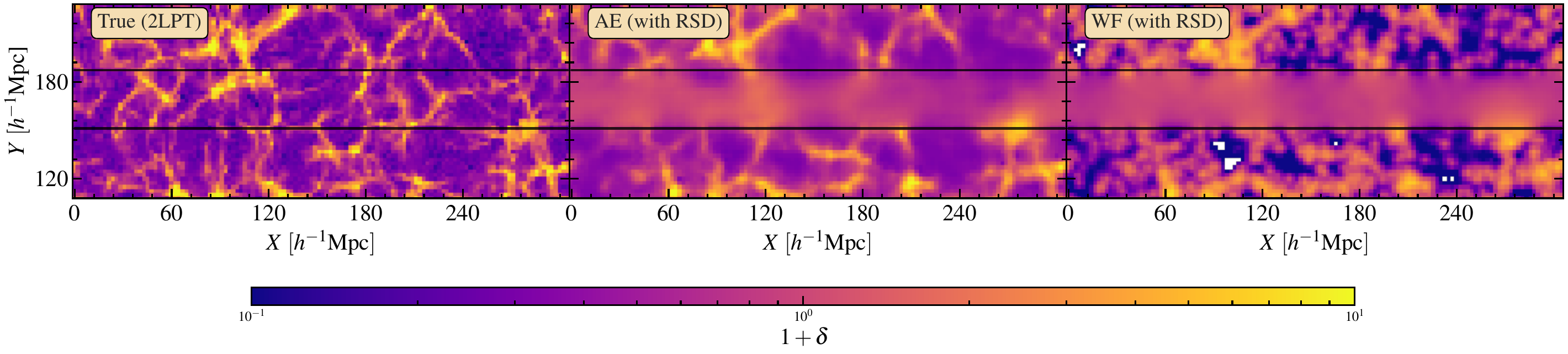} \vskip -0.2cm
    \caption{Reconstructions {of the} 2LPT {density field} with a slab-shaped mask of $35 \, \hmpc$ in thickness, in the $Y-Z$ plane, as marked by the two horizontal lines. The left panel is the true field, the center and right panels show the reconstructions using {the} \Ae\ and WF, respectively, {with RSD}.}
    \label{fig:2LPT_density_fields_masked}
\end{figure*}

We  mask the observed redshift-space density in a region spanning the whole $X-Z$ plane with a slab of thickness of $\Delta Y \approx 35 \,\hmpc$ (15 bins).
In \cref{fig:2LPT_density_fields_masked} we show the true and reconstructed density fields (same realization as for the unmasked case in \cref{fig:2LPT_density_fields}). We plot only the  $Y$ range close to the mask as indicated  by horizontal black lines. Outside of the mask, both \Ae\ and WF reconstructions closely match their respective unmasked counterparts. Inside the mask, however the reconstruction quality  deteriorates, as both the \Ae\ and WF fields tend to approach the mean value of {the} density contrast, $\delta=0$. Although, the  \Ae\ 
displays more pronounced  structures in the masked region, some of these structures are absent from the true underlying field. For example, at $X \approx 120 \, \hmpc$ the \Ae\ predicts an overdense region spanning the whole mask, but which does not exist in the true field (there is only a true overdensity in the upper part of the mask).

\begin{figure}
    \centering
    \includegraphics[width=\columnwidth]{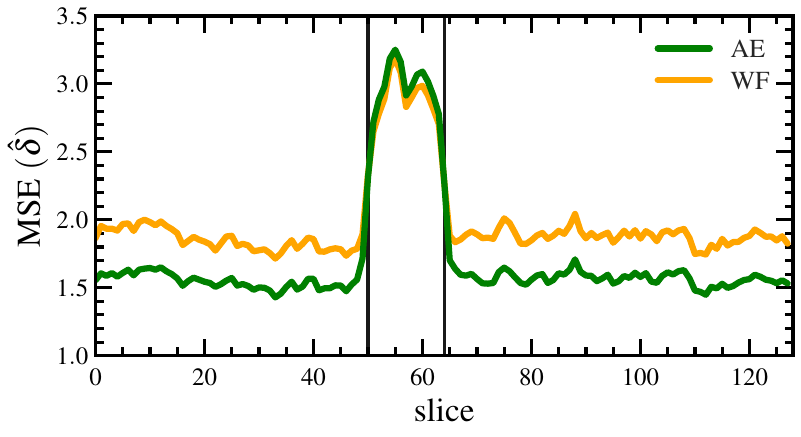}
    \vskip -0.05cm
    \caption{ The MSE per slice averaged over the entire test set of the 2LPT density fields reconstructed from a 
    masked input density, as described in the text. The gray lines indicate the masked region.}
    \label{fig:my_label}
\end{figure}

\Cref{fig:2LPT_density_fields_masked} plots the MSEs of the reconstructed fields in individual $X-Z$ slices parallel to the masked region. Outside the mask (marked by vertical black lines), the MSEs for both reconstructions match the unmasked case. At the mask boundaries both MSEs rise sharply and approach the (cosmic) variance of the underlying density (cf.~\cref{tab:density_statistics}) a few Mpc into the mask. Due to the short correlation length of the density field, the \Ae\ hence does not reduce the MSE within the mask compared to the WF. In fact, the WF MSE in the mask is slightly smaller than the \Ae\ MSE. The difference is practically negligible, though, and we expect it {to} vanish when increasing the number of training samples.

\begin{figure*}
    \includegraphics[width=\textwidth]{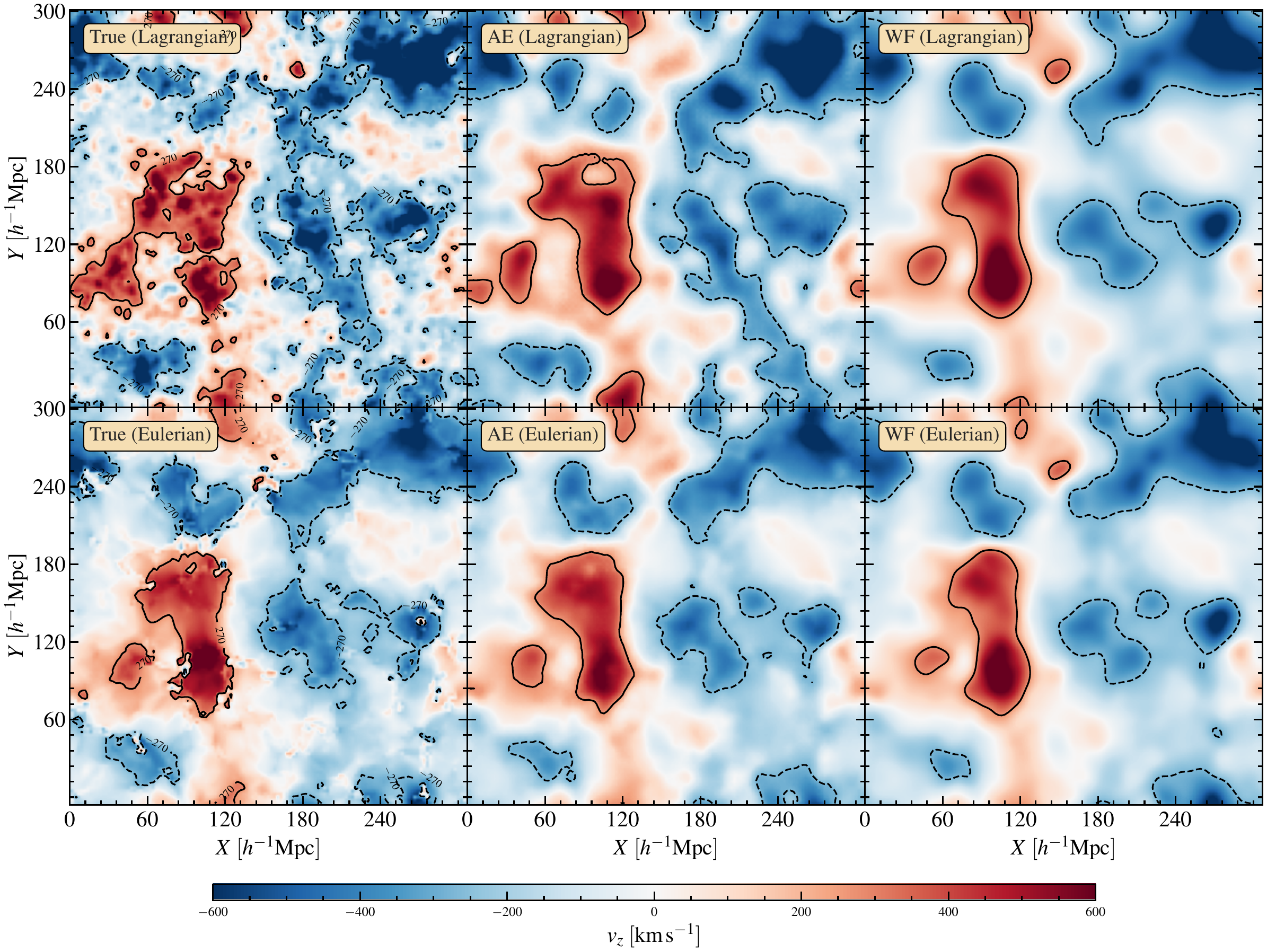}
    \vskip -0.07cm
    \caption{Line-of-sight ($z$-direction) velocity fields through the $X-Y$ plane reconstructed from
    the observed density field {with RSD} plotted in \cref{fig:2LPT_density_fields}.
Top and bottom panels  correspond to  Lagrangian and Eulerian space.
 The left panels display the true $v_z$, while the middle and right panels are  the corresponding \Ae\ and WF reconstructions.  Values 
 $-270$ and $+270 \kms$ are designated by the dashed and solid contours, respectively.}
    \label{fig:2LPT_velocity_fields}
\end{figure*}

\begin{figure*}
  	\centering
  	\begin{tabular}{rr}
  	\textbf{Lagrangian $v_{z}$} \hspace{0.34\columnwidth} &
  	\textbf{Eulerian $v_{z}$} \hspace{0.37\columnwidth} \\
  	\imagetop{\includegraphics[width=0.92\columnwidth]{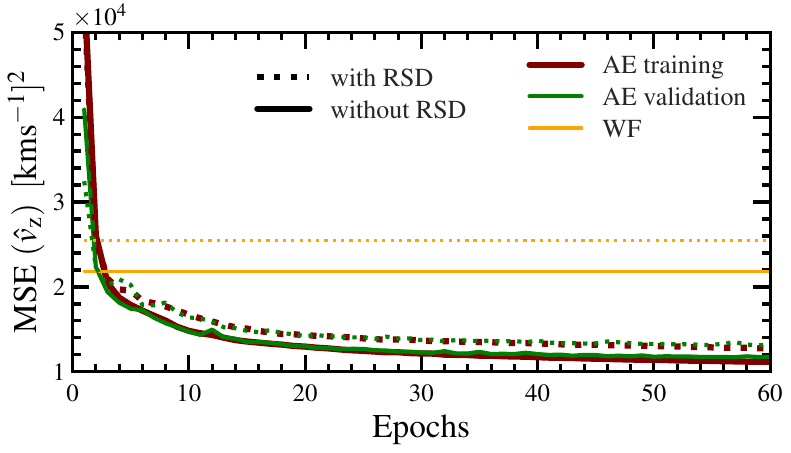} \hspace*{0.006\columnwidth}} &	
\imagetop{\includegraphics[width=0.94\columnwidth]{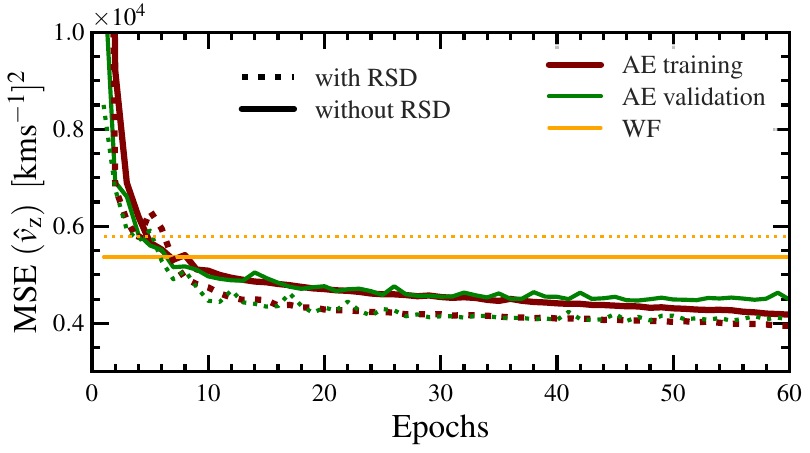} \hspace*{0.007\columnwidth}} \\
\imagetop{\includegraphics[width=0.98\columnwidth]{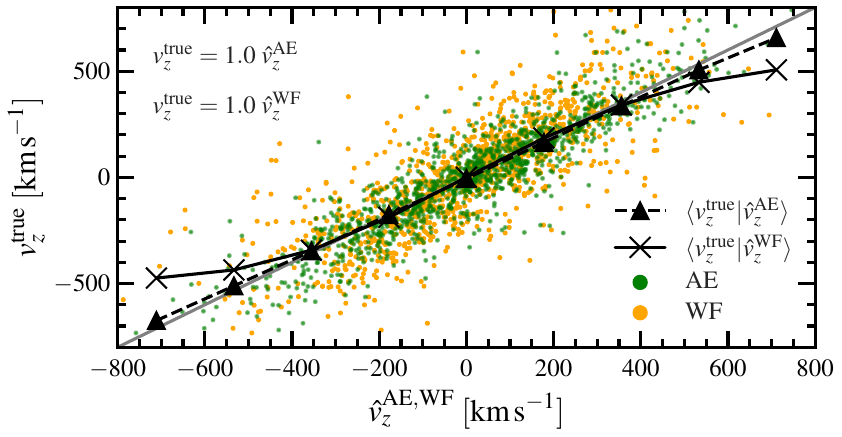} \hspace*{0.0005\columnwidth}} &
  	\imagetop{\includegraphics[width=0.98\columnwidth]{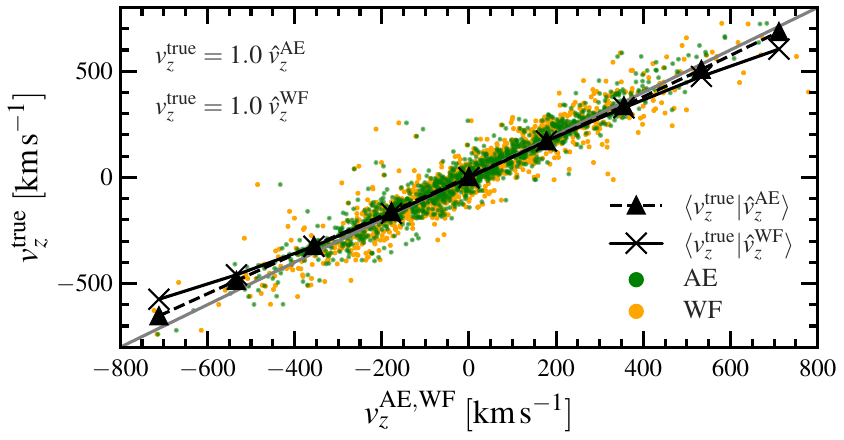} \hspace*{0.001\columnwidth}} \\
  	\imagetop{\includegraphics[width=0.98\columnwidth]{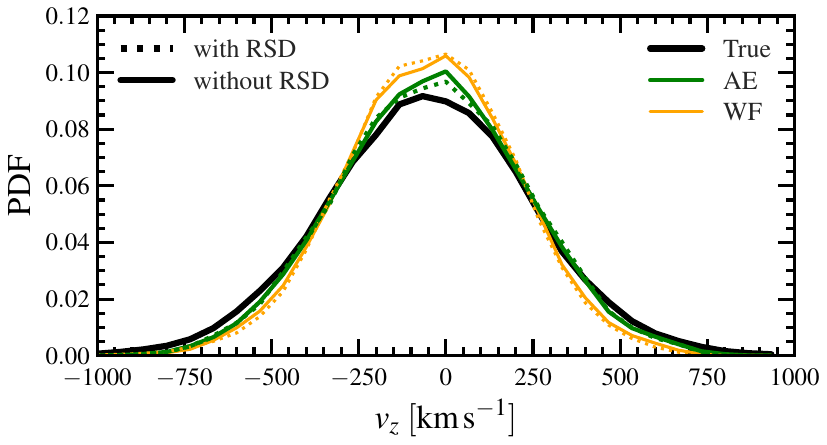}} &
  	\imagetop{\includegraphics[width=0.96\columnwidth]{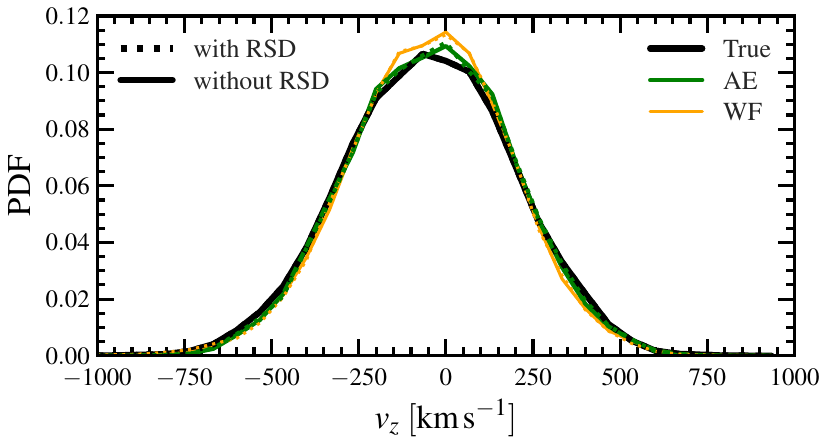}}
  	\end{tabular}
   \vskip -0.05cm
	\caption{
	\textit{Top panels:} the MSE versus {the training} epoch for various reconstructions as indicated in the panels. 
  \textit{Middle panels:} 
 Scatter plot of true versus reconstructed $v_z$ velocities at randomly selected grid cells for {the} one realization {shown in \cref{fig:2LPT_density_fields}}.  {The slopes} of linear regressions are indicated in the figure. The black crosses and triangles represent the mean $v_z^{\mathrm{true}}$ in a given bin of $\hat v_z^{\mathrm{\wf}}$ and $\hat v_z^{\mathrm{AE}}$ respectively, over all grid cells. 
  \textit{Bottom panels:} PDFs of true and reconstructed $v_z$. 
  The Left and right columns correspond to Lagrangian and Eulerian $v_z$. 
  } 
	\label{fig:2LPT_velocity_statistics}
\end{figure*}

\subsection{2LPT velocity reconstruction}
\label{sec:results:2LPT_velocity}
We turn now to the reconstruction of the underlying velocity field (in Lagrangian and Eulerian space) from the observed density. Of particular interest is the line-of-sight velocity component, as it can be directly compared with the line-of-sight peculiar velocities inferred from galaxy distance catalogs. In our 2LPT data set, this corresponds to the $z$-component, $v_z$, of the velocity field.

\Cref{fig:2LPT_velocity_fields} plots the $v_z$ fields corresponding to the density field in \cref{fig:2LPT_density_fields} in the same slice through the $X-Y$ plane. True and reconstructed velocity  fields, both in Lagrangian and Eulerian space, are plotted as indicated in the figure. We only show reconstructions from the observed density including RSD, as they are visually similar to reconstructions without RSD. For the Lagrangian $v_z$ (top panels), both \Ae\ and WF fields capture the large-scale fluctuations in the true $v_z$. Despite the similarities, intermediate-scale structures are more pronounced in the \Ae\ reconstruction, as outlined by the contours designating the values $-270\,\kms$ (dashed) and $+270\,\kms$ (solid).

The Eulerian $v_z$ is the mean velocity of particles lying in the same Eulerian grid cell but having substantially different Lagrangian positions. As a result, the true Eulerian field (bottom) appears smoother than its  Lagrangian counterpart. Overall, both reconstructions match the underlying field more closely than in the Lagrangian case, with the \Ae\ still faring better than the WF.

To better assess the quality of $v_z$ reconstructions, we compare in {the top panels of} \cref{fig:2LPT_velocity_statistics} {the MSE loss} versus  the training epoch.  For both Lagrangian (left) and Eulerian (right) $v_z$, the MSE of \Ae\  drops below that of WF after only a few epochs and converges to a minimum after  $30$\,--\,$40$ epochs. 

For the Eulerian case, we notice a slight over-fitting as indicated by the mild departure between the training and validation MSEs beyond epoch $40$. The validation MSE remains nearly constant while the training MSE drops by  less than $10\%$ for over $20$ epochs.
In agreement with the overall visual impressions drawn {from} \cref{fig:2LPT_velocity_fields}, the MSE is much smaller for the Eulerian $v_z$. Further,  the relative reduction of the MSE for \Ae\ compared to WF is larger for the Lagrangian $v_z$. Precise MSE values can be found in \cref{tab:velocity_statistics}. For both Lagrangian and Eulerian $v_z$, the relative difference between the MSEs with and without RSD is smaller than it is for the density  (cf.~\cref{fig:2LPT_density_statistics}). While in most cases the MSE with RSD is larger than the one without RSD, this is reversed for the \Ae\ MSE of the Eulerian velocity. It is unclear why this one case differs from the rest, but the effect is small nonetheless.

The middle panels of \cref{fig:2LPT_velocity_statistics} are  scatter plots of the true versus reconstructed $v_z$, for a random subset of grid cells in one realization with RSD (similar results are found  without RSD). In both panels, the scatter is visually smaller in the \Ae\ reconstruction compared to the WF. In both velocity spaces, {a} linear regression slope very close to unity is found for the \Ae\ as well as the WF, while the the inverse regression of reconstructed versus true $v_z$ yields slopes significantly  smaller than unity.
Also plotted are {the} conditional true  vs reconstructed $v_z$ as triangles (for \Ae) and crosses (WF). As for the density reconstruction, the triangles are very close to the diagonal. This is in agreement with the theoretical expectation for the mean posterior estimate discussed in \cref{sec:methodology:estimator_properties}.
At large absolute velocities, the WF, as indicated by the crosses, underestimates the true $v_z$. This is expected since generally the WF estimate is not equal to the mean true $v_z$.

The bottom row in the figure presents the PDFs of the various fields, computed  from the test samples. For both the Lagrangian (left) and Eulerian (right) $v_z$, the inclusion of the RSD in the observed density has little {effect} on the PDF, as confirmed by comparing the dashed and solid curves for each case. For the Lagrangian $v_z$, the deviation of the reconstructed PDF from the true one is significant, with the \Ae\ PDFs being clearly closer to the true PDF than the WF. In accordance with the visual impression from \cref{fig:2LPT_velocity_fields}, the PDFs of the reconstructed Eulerian $v_z$ are closer to the true PDF.

\subsubsection{Effect of masking}
\label{sec:results:2LPT_velocity:mask}
We now investigate how masking part of the observed density field affects the quality of the velocity reconstruction. The same mask shape as for the density reconstruction is considered, described in \cref{sec:results:2LPT_density:mask}. We focus only on the Lagrangian velocity reconstruction with RSD, since the masking effects are qualitatively the same in Eulerian space for both with and without RSD.

\begin{figure*}
\includegraphics[width=\textwidth]{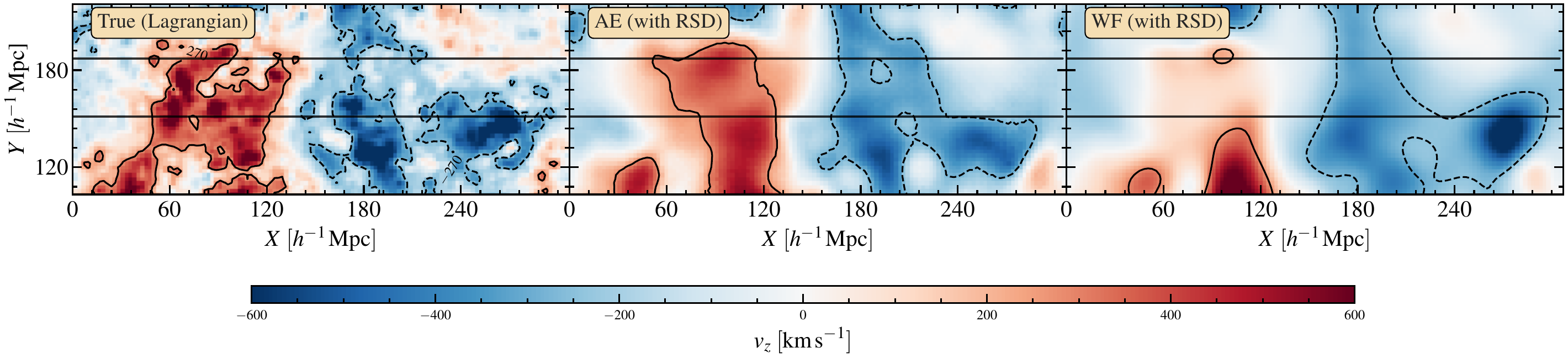}
\vskip -0.2cm    \caption{ The  line-of-sight Lagrangian $v_z$ from a masked observed density field with RSD. The slab-shaped mask  is marked by the two horizontal lines. The panels from left to right show the true velocity field, \Ae\ and WF reconstructions, respectively.
    }
    \label{fig:2LPT_velocity_fields_masked}
\end{figure*}

\Cref{fig:2LPT_velocity_fields_masked} shows the true and reconstructed Lagrangian $v_z$ field (same realization as for the unmasked case in the top panels of \cref{fig:2LPT_velocity_fields}) in the $Y$ range close to the mask (marked by horizontal black lines). At larger distances from the mask, the reconstructions match those in the unmasked case. Inside the mask and in its nearest vicinity, the \Ae\ reconstruction  captures less of the underlying small-scale structure compared to the unmasked case, but it still manages to recover the dominant large-scale features. The WF reconstruction clearly recovers fewer of these larger-scale features and decays more quickly towards the {zero} mean {velocity} within the masked region.

\begin{figure}
    \centering
    \includegraphics[width=\columnwidth]{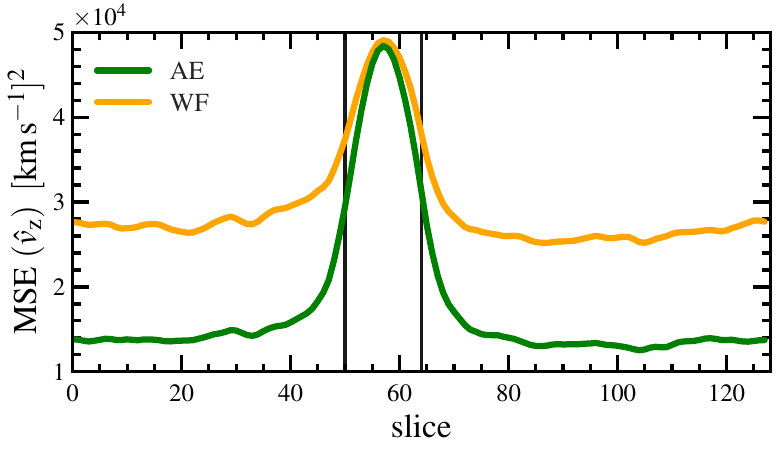}
    \vskip -0.05cm
    \caption{MSE of the reconstructed line-of-sight Lagrangian velocity  from {the} masked  {observed} density field including RSD.
    }
    \label{fig:2LPT_velocity_statistics_masked}
\end{figure}

In \cref{fig:2LPT_velocity_statistics_masked} we quantify the visual impression  by plotting the MSE in individual $X-Z$ slices parallel to the masked region. Far from the mask, the MSE per slice agrees with the overall unmasked MSE shown in the top left panel of \cref{fig:2LPT_velocity_statistics} for both \Ae\ and WF. When approaching the mask, the MSEs of both estimators quickly increase. This increase starts at much larger distances from the mask ($\sim 40 \, \hmpc$) than for the density reconstruction in \cref{fig:2LPT_density_fields_masked}, due to the longer correlation length of the velocity field. For all slices, the \Ae\ MSE stays below the WF MSE, but the relative difference between both decreases towards the center of the mask, where they are nearly identical, reaching  $\sim 65\%$ of the cosmic variance (cf.~\cref{tab:velocity_statistics}). The more accurate \Ae\ reconstruction in particular around the borders of the mask explains why the \Ae\ is capable of recovering the underlying structures further into the masked region than the WF.

\begin{table*}
    \centering
    \caption{Statistical properties of the true and reconstructed velocity fields{, computed} from the test set.}
    \label{tab:velocity_statistics}
	\begin{tabular}{ @{}llccccc@{}} 
       \toprule
	&	\textbf{Coordinates}  & \textbf{Field} & \textbf{Variance} & \textbf{MSE} &
		\textbf{Regression slope} &  \textbf{Regression slope} \\
			&	  & $ v_z^{\mathrm{field}} $ & $ \langle (v_z^ {\mathrm{field}} )^{2}\rangle$ & $ \langle (v_z^{\mathrm{true}} - v_z^ {\mathrm{field}}) ^ {2} \rangle$ & \textbf{ $ v_z^{\mathrm{true}}$} {vs} {$v_z^\mathrm{field}$ } &  \textbf{$v_z^\mathrm{field}$}  {vs}  {$v_z^{\mathrm{true}}$} \\
			& & & $[100 \, \kms]^2$&$ [100 \, \kms]^2 $ & & \\
    		\toprule	
    		\multirow{6}{*}{\textbf{2LPT without RSD}} 
	 &  &True & $  7.8  $ & $ -$ 
	     & $-$ & $-$\\
&Lagrangian	&Autoencoder & $ 6.9  $ & $ 1.2 $ &
	$ 1.0$ & $0.84 $\\ 
    	 & 	& Wiener filter & 
    		$ 6.0  $ & $ 2.2  $ 
    		& $1.0 $ & $0.71 $ \\
    		\cmidrule{2-7}
    		&  &True & $5.6   $ & $ -$ & $ -$ & $ -$\\
    	&	Eulerian & Autoencoder & $ 5.3 $ & 
	$ 0.45 $ & $  0.99  $ & $ 0.93$\\ 
    	&  	& Wiener filter & $5.0 $&
    		$ 0.54$
    		& $  1.0 $ & $0.89$ \\
    		\midrule
    	\multirow{6}{*}{\textbf{2LPT with RSD}} 
    	&Lagrangian	& Autoencoder & $ 6.2  $ & 
    		$ 1.3 $ & $1.0$ & $ 0.81  $  \\
      & & Wiener filter &$ 5.4  $ & $ 2.6 $  & $0.99 $ & $ 0.68 $\\
    		\cmidrule{2-7}
    &		Eulerian		& Autoencoder & $ 6.2 $ & $ 0.42 $  & 
    		$ 1.0$ & $0.99   $  \\
    & 		& Wiener filter & $5.4 $ & $ 0.59 $
    		& $0.99 $ & $ 0.89$ \\
    	\bottomrule
    \end{tabular}
\end{table*}

\section{Dependence of the MSE on the average 
 number density}
\label{sec:nbar}
   
The MSE is a figure of merit for the quality of the reconstruction. Thus it is useful to explore its dependence on the mean number density, $\bar n$, of galaxies in the sample.  
For the most straightforward estimate of the density, $\hat \delta=\delta^\textrm{obs}$, the MSE is $(\bar n \, \Delta V)^{-1}$ where $\Delta V$ is the volume per  grid cell. 
For the WF estimator (cf.~\cref{sec:WF_computation}),
the ${\bar n}^{-1}$ scaling is retrieved only for large $\bar n$. 

\begin{figure*}
    \centering
    \includegraphics[width=\columnwidth]{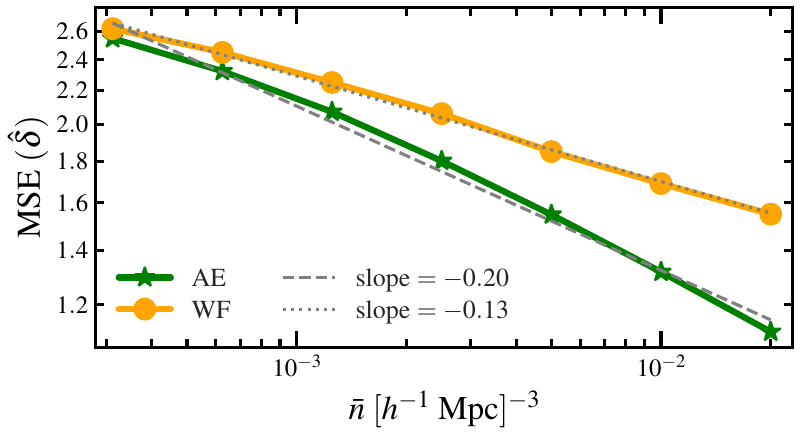}
    \hfill
    \includegraphics[width=\columnwidth]{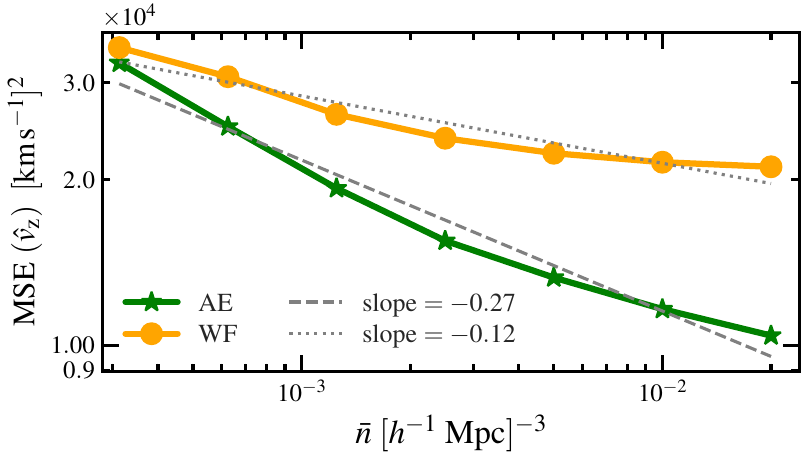}
    \vskip -0.02cm
    \caption{
    Dependence of {the} MSE on the mean observed galaxy number density. \textit{Left panel}{: MSE of the density reconstruction.} The green solid line is for the \Ae\, which has a consistently lower MSE than {the} WF (amber line). The straight grey dashed and dotted lines show the average slopes of the AE and WF data points, respectively. The relative difference between the two increases with an increase in galaxy number density. For reference, the total range of mean densities shown here approximately corresponds to a distance range of $30 \, \hmpc$ to $170 \, \hmpc$ in the 2MASS Redshift Survey. The two leftmost points also correspond to the galaxy number densities expected to be observed roughly around redshifts $z \sim 1.1$ and $z \sim 1.5$ in the Euclid spectroscopic survey \citep{euclid_collaboration_euclid_2020}. \textit{Right panel}: The same but for the velocity reconstructions.
    }
    \label{fig:nbar}
\end{figure*}

In \cref{fig:nbar} we plot the MSE versus $\bar n$ for the AE- and WF-reconstructed density (left panel) and velocity (right) fields. The plotted values, $3.125 \times 10^{-4} \, \hmpcdens \leq \bar n \leq 2 \times 10^{-2} \, \hmpcdens$,  roughly cover the range of typical observed galaxy densities in spectroscopic galaxy surveys. The high-$\bar n$ points match the inner regions of the 2MASS Redshift Survey (2MRS) \citep{huchra_2mass_2012}, while the low-$\bar n$ points correspond to both the outer regions of 2MRS and the expected number density range from the Euclid spectroscopic survey \citep{euclid_collaboration_euclid_2020}. For the density field, both reconstructions yield a substantially shallower scaling $\textrm{MSE}\sim {\bar n}^\alpha$ than ${\bar n}^{-1}$.  For the WF the quality of the reconstruction improves very little 
with increasing  number density, $\alpha=-0.13$ on average, whereas the \Ae\ reconstruction improves more strongly, $\alpha=-0.20$.
For the velocity field reconstruction, the MSE shows on average a similar dependence on $\bar n$ for the WF, $\alpha=-0.12$, and a steeper dependence for the AE, $\alpha-0.27$.

We have also tested the dependence of the density  PDF on $\bar n$. The main impact  is on the  minimum density cutoff seen in the \Ae\ reconstruction (e.g.~top-right and bottom-left panels in 
\cref{fig:2LPT_density_statistics}) and the high-density tail. As $\bar n$ is increased, the cutoff is lowered  and the tail approaches the true PDF.

\section{Discussion and conclusions}
\label{sec:discussion}
The \Ae\ is designed to yield (for a sufficiently large training set) the mean of the true density/velocity that is consistent with the observations. As a consequence, the \Ae\ is capable of capturing  nonlinear  effects in its reconstructions. In contrast, as a linear operation on the observed field, the WF agrees with the \Ae\ \emph{only} for Gaussian fields.

The \Ae\ and WF yield comparable results for the reconstruction of the density field from noisy data but there are important differences. The MSE is lower by 15\%  for \Ae\ 
and the density does not reach negative values as in the WF. In fact, the \Ae\  reconstruction is characterized by a density floor that depends on the noise level.  This is the result of the Poisson scatter present in the input (observed) density field estimated directly from the discrete distribution of points in the mocks.

We also tested, but have not presented the results in the main body of the paper, our \Ae\ 
applied to 2LPT fields in a $128^3$ cubic box with a larger physical size of $1000 \,\hmpc $ and a lower galaxy density of $5\times10^{-4}\, \hmpcdens$. This approximately corresponds to the expected galaxy density around a redshift of $z \sim 1.3$ in the Euclid spectroscopic survey \citep{euclid_collaboration_euclid_2020}.
Despite the larger, more linear scales probed by this box, the reduction of the MSE($\hat{\delta})$ for the AE (0.54) relative to the WF  (0.61) is comparable to the relative reduction found for the smaller $300 \,\hmpc $  box. The skewness of the true density field (4.67) is also recovered more accurately by the AE (4.74) than by the WF (1.81). Furthermore, the training converged faster with fewer   training samples, even as low as 30 samples.
This motivates  potential applications to large galaxy surveys such as Euclid.

We have demonstrated that an AE is superior to the
WF for the purpose of reconstructing peculiar velocity fields from the distribution  of mass tracers in redshift space. The reason for this is the nonlinear and nonlocal 
nature of the density-velocity relation. The WF is based on linear gravitational instability theory and {thus} incapable of mapping nonlinear features in the density-velocity relation.  The  \Ae\ can both capture nonlinearities as well as any correlations in the relation. Unfortunately, since the modelling of \Ae\ is done via an intricate combination of weights, the mapping itself cannot be inferred in a closed form and therefore is rendered  useless for deriving any insight regarding nonlinear gravitational dynamics.

We needed more samples to train the \Ae\ for {the} velocity reconstructions compared to {the} density.  
This could be due to the longer coherence length of velocity fields, implying  {that large-}scale modes  contribute more to the MSE  for {the} velocity compared to {the} density. Thus, to compensate for the fewer large-scale modes {per sample}, more  training samples are needed. 

We demonstrated that the \Ae\ can handle partially masked observations, despite the fact that each individual \Ae\ layer applies a translationally invariant convolution or pooling operation. This is in line with previous findings that sufficiently deep convolutional NNs can learn features that break translational invariance, such as a mask, by exploiting boundary effects \citep{semih_kayhan_translation_2020}. For the density reconstruction, the MSEs of \Ae\ and WF inside the masked region are comparable. For the velocity reconstruction, however, the \Ae\ MSE stays below the WF MSE for more than $15 \, \hmpc$ into the mask. This difference is due to the larger correlation length of the velocity field compared to the density field.

It should be emphasized that a NN can only  faithfully capture the information encoded  in the mock training data. 
Since  the \Ae\   relies on training sets and otherwise is generally unaided by theoretical assumptions, it is particularly important to generate as realistic mock data as possible. 
A caveat  of the current work is that the mocks are 
based on 2LPT, which is reasonable for modelling dynamics on mildly nonlinear scales, but does not allow the identification of halos and hence galaxies inside the simulation. Therefore, the observed galaxy number counts in our mocks are obtained by simply selecting a random subset of dark matter particles, which produces realistic galaxy number densities including Poisson noise, but does not account for galaxy bias. Although galaxy bias will affect the MSE values of the AE and the WF, we do not expect it to change  the conclusions of the present paper. In fact,  realistic galaxy biasing  adds complexity to  the mapping between observed and underlying fields, which further  motivates the \Ae\ as a reconstruction tool. The employed simplified mocks are sufficient for our purposes here of preliminary assessing the  \Ae\ for the reconstruction of 3D fields and 
clarifying the connection to the mean posterior and WF estimators.

The next step will then be to train the \Ae\ on realistic mock galaxy catalogs extracted from N-body simulations, including galaxy bias as well as observational selection effects 
in flux/magnitude-limited redshift surveys. This will allow us to quantify the achievable reconstruction accuracy of the AE for actual surveys more accurately.
Fortunately, as we have shown, the \Ae\ reconstruction does not require an
excessively large number of training samples. Thus,  creating a sufficient number of mock catalogs based on 
simulations that incorporate more accurate dynamics than 2LPT 
should be entirely feasible \citep[e.g.][]{Villaescusa-Navarro_Camels_2022, Maksimova_Abacus_2021}.

In the present work, we have focused on the reconstruction of the velocity from an observed density field in redshift space. 
We have not considered the inverse problem 
of  inferring the   density field associated with measurements of peculiar velocities of galaxies.
{So far, the} inference of the DM density field has been {achieved} either via direct smoothing that takes into account the sparseness of data such as in the POTENT method \citep{dek93}, via {the} WF \citep{zaroubi_wiener_1999,courtois_three-dimensional_2012}, or via hierarchical Bayesian models \citep{graziani_peculiar_2019,valade_hamiltonian_2022}.
These techniques  relied either on linear theory or on simulation-calibrated  mildly nonlinear relations between velocity and density
\citep[e.g.][]{nussetal91}.

Putting the issue of observational biases aside, e.g.~Malmquist bias \citep[e.g.][]{Lynden-Bell1988}, the reconstruction of  the density field from velocity measurements is substantially more challenging than the
problem of {reconstructing the} velocity from {an observed} density. The relatively small number of measurements in peculiar velocity catalogs and their highly  sparse spatial coverage prevent a straightforward application of convolutions in the NN. One way to proceed is to smooth  (e.g.~a tensor smoothing {\`a} la POTENT)  the measurement to provide a radial velocity on a regular grid. 
A training set  of mock velocity data 
smoothed similarly as the real data is then  suitable 
for  a {convolutional} NN. Another way is to explore NN architectures designed specifically for sparse input data, e.g.~graph NNs \citep[e.g.][]{villanueva-domingo_learning_2022,makinen_cosmic_2022} or NNs designed to analyse point cloud data \citep[e.g.][]{anagnostidis_cosmology_2022}.

In future work, we will  apply our \Ae\ reconstruction to the 2MASS Redshift Survey \citep{huchra_2mass_2012,macri_2mass_2019} by training  on mock catalogs mimicking the selection criteria of the survey. 
Furthermore, in the coming decade several large-scale sky surveys such as DESI \citep{desi_collaboration_desi_2016}, Euclid \citep{euclid_collaboration_euclid_2018}, LSST \citep{lsst_science_collaboration_lsst_2009}, 4MOST \citep{4mostCollaboration2012}, SPHEREx \citep{Spherex_cosmology_2014} and others will map a very large volume of the Universe by cataloguing tens of millions to billions of galaxies. We believe that reconstructing the large-scale density and velocity fields from this large data set, to extract the cosmological parameters, will benefit from {employing} NN methods  over the traditional techniques. 

\section*{Acknowledgements}
We thank Lihi Zelnik-Manor, Ofer Lahav and Gal Ness for stimulating discussions. This research is supported by the Israel Science Foundation (ISF) grant \#$893/22$ and the Asher Space Research Institute.

\subsection*{Data Availability}
The data underlying this article will be shared on reasonable request to the corresponding author. 

\bibliographystyle{mnras}
\bibliography{bibliography_PGV,bibliography_RL,bibliography_AN}



\appendix

\section{Generating 2LPT training {data}}
\label{sec:2LPT_data_generation}
We have used  2LPT-generated density and velocity fields as training data since they are inexpensive to generate but display all features required for our proof-of-concept analysis: non-Gaussian density distribution, nonlinear relation between density and velocity, and visual similarity to the actual cosmological fields. A comprehensive overview of the relevant 2LPT equations can, for example, be found in \citet{scoccimarro_transients_1998}. In the following, we give a concise summary of our implementation.

We generate the fields in a periodic cubic box on a $128 \times 128 \times 128$ grid. The Eulerian density is obtained by displacing a uniform Lagrangian distribution of $512^3$ particles according to
\begin{equation}
    \vect{r}(\vect{q}) - \vect{q} = - D_1 \, \nabla_q \phi^{(1)}(\vect{q}) + D_2 \, \nabla_q \phi^{(2)}(\vect{q}) \,,
\end{equation}
with the usual linear growth factor $D_1$ \citep{Peeb80}, the second-order growth factor $D_2 \approx -\frac{3}{7} \, D_1^2$, and the two Lagrangian potentials following, respectively,  the Poisson equations
\begin{align}
    \nabla_q^2 \phi^{(1)}(\vect{q}) &= \delta(\vect{q}) \,, \\
    \nabla_q^2 \phi^{(2)}(\vect{q}) &= \! \sum_{i>j} \Bigl[ \nabla_{q_i} \! \nabla_{q_i} \phi(\vect{q}) \, \nabla_{q_j} \! \nabla_{q_j} \phi(\vect{q}) \! - \! \bigl[\nabla_{q_i} \! \nabla_{q_j} \phi(\vect{q})\bigr]^2 \Bigr] .
\end{align}
Here, the Lagrangian-space density contrast $\delta(\vect{q})$ is a correlated random Gaussian field following the linear power spectrum computed with CLASS \citep{lesgourgues_cosmic_2011}.  We additionally smooth the power spectrum with a Gaussian window of width $1 \, \hmpc$, to suppress stream-crossing on the smallest scales resolved in this box. We set $\Omega_\mathrm{m} = 0.3$ and evolve to $D_1 = 1$, to obtain a fluctuation amplitude $\sigma_8$ close to the actual value at redshift $z=0$.

The Poisson equations are efficiently computed via FFT operations. Afterwards, the Eulerian-space density field is computed from the particle counts, where each displaced particle is assigned to its nearest Eulerian grid cell. To obtain the associated (real-space) observed density field, we introduce Poisson noise by selecting a random subset of $N_\text{obs} = \bar{n} V$ particles, where $\bar{n}$ is the desired mean observed galaxy density and $V$ is the box volume.

In our training data, we consider two different velocity fields as targets to be reconstructed: (i) the Lagrangian-space velocity field,
\begin{equation}
    \vect{v}(\vect{q}) = - D_1 f_1 H \, \nabla_q \phi^{(1)}(\vect{q}) + D_2 f_2 H \, \nabla_q \phi^{(2)}(\vect{q}) \,,
\end{equation}
with the usual linear growth rate $f_1 \approx \Omega_\mathrm{m}^{5/9}$, the second-order growth rate $f_2 \approx 2 \, \Omega_\mathrm{m}^{6/11}$ and the Hubble constant $H = 100 \, h \, \kms \, \mathrm{Mpc}^{-1}${;} (ii) the Eulerian-space velocity field $\vect{v}(\vect{x})$ computed by averaging the velocities of all particles associated with an Eulerian grid cell.

For the reconstructions from redshift space, we additionally generate observed redshift-space density fields by shifting the observed galaxies along the $z$-axis by $v_z(\vect{q}) / H$ before assigning them to the nearest Eulerian grid cell. We are thus treating the $z$-axis of the box as our line-of-sight direction.

\section{Computation of Wiener filter}
\label{sec:WF_computation}
The reference Wiener filter (WF) reconstructions are obtained in two different ways.

In the case of unmasked data, the observed and reconstructed fields are translationally invariant. Thus, we can diagonalize the WF given in \cref{eq:WF_weights} by Fourier-transforming it, yielding
\begin{align}
    \hat{\delta}^{(\mathrm{WF})}(\vect{k}) &= \frac{P_{\delta \delta'}(k)}{P_{\delta' \! \delta'}(k) + \bar{n}^{-1}} \, \delta^{\prime(\mathrm{obs})}(\vect{k}) \,,
    \label{eq:WF_density_unmasked} \\
    \hat{\vect{v}}^{(\mathrm{WF})}(\vect{k}) &= \frac{\mathrm{i} \vect{k}}{k^2} \, \frac{P_{\theta \delta'}(k)}{P_{\delta' \! \delta'}(k) + \bar{n}^{-1}} \, \delta^{\prime(\mathrm{obs})}(\vect{k}) \,.
    \label{eq:WF_velocity_unmasked}
\end{align}
Here, $\delta'$ denotes the density contrast in the space (real or redshift) in which the observed density contrast is given, $\theta = \nabla \cdot \vect{v}$ is the (Lagrangian or Eulerian) velocity divergence, and $P_{XY}$ is the nonlinear power spectrum between two fields $X, Y \in \{\delta, \delta', \theta\}$. The various power spectra are computed from a set of ten 2LPT density and velocity field realizations.

In the case of  masked data, the translational  invariance is broken and we resort to an iterative scheme for performing the multiplication with the inverse covariance matrix of the observed density contrast in \cref{eq:WF_weights}. For this we follow the conjugate gradient descent method described in \citet{kitaura_bayesian_2008}. We also confirmed that in the unmasked case the iterative approach agrees with the direct application of the WF in Fourier space.



\bsp 
\label{lastpage}

\end{document}